\documentclass{valence} 

\usepackage{amsmath,amsfonts,bm}

\def\eqref#1{equation~\ref{#1}}

\def\1{\bm{1}}

\DeclareMathAlphabet{\mathsfit}{\encodingdefault}{\sfdefault}{m}{sl}
\SetMathAlphabet{\mathsfit}{bold}{\encodingdefault}{\sfdefault}{bx}{n}

\usepackage[T1]{fontenc}
\usepackage{charter}

\usepackage{graphicx} 
\usepackage{booktabs}
\usepackage{hyperref}
\usepackage{makecell}
\usepackage{amsmath}
\usepackage{amssymb}
\usepackage{natbib}
\usepackage{float}
\usepackage{amsmath}
\usepackage{xfrac}
\usepackage{tikz}
\usetikzlibrary{arrows.meta, positioning, calc, backgrounds, fit, decorations.pathreplacing, shadows}

\hypersetup{colorlinks=true,allcolors=[rgb]{0.,0.5,0.5}}

\title{AquaGen: Scaling generative models to molecular dynamics precision on thousands of atoms}
\author{Emmanuel Bengio$^{^\star 1,2}$, Sanjeev Raja$^{^\star 1,2,\dagger}$, Yui Tik Pang$^{1,2}$, Kerstin Klaeser$^{1,2}$, Cristian Gabellini$^{1,2}$, Nikhil Shenoy$^{1,2}$, Francesco Di Giovanni$^{1,2}$, Prudencio Tossou$^{1,2}$}

\affiliation[1]{Valence Labs}
\affiliation[2]{Recursion}
\affiliation[\star]{Equal Contribution}

\footnotetext[1]{UC Berkeley; work done during an internship at Valence Labs}

\abstract{We present \emph{AquaGen}, the first all-atom, explicit solvent, periodic-boundary-condition-aware generative model that produces molecular configurations from the Boltzmann distribution at a fraction of the cost of molecular dynamics (MD).  This is in contrast with existing generative models that remove degrees of freedom by operating on coarse-grained, vacuum, or implicit solvent systems. Operating at this resolution allows for post-processing through force field energy evaluations and MD simulations, and enables the prediction of relevant properties in a \emph{gray-box} manner (as ensemble averages of potential energy evaluations over generated samples). We demonstrate the utility of this paradigm on absolute hydration free energy (AHFE), producing estimates 4-10x faster and with comparable accuracy to standard GPU-based MD. By generating uncorrelated samples from alchemical Boltzmann distributions, we create more accurate, interpretable, and refinable ensemble predictions with calibrated uncertainty estimates, unlike regression methods which are entirely \emph{black-box} predictors. Our approach also yields predictable benefits from increasing train- and test-time compute, realized by scaling model size and generating more samples, respectively.  We believe that this approach demonstrates the utility of high-resolution ensemble generation for free energy estimation, with future potential to replace MD in tasks such as the prediction of lipophilicity, membrane permeability, or absolute binding free energy (ABFE) -- whose grounding and interpretability may be critical for the development of new drugs and materials.}

\begin{document}

\maketitle

\newif\ifvtwofig
\vtwofigtrue

\begin{figure}[ht]
    \centering
    \newcommand{\figoneinlined}{hi}

    \makebox[0pt][l]{\hspace{-0.49\linewidth}\small(a)}
    \input{fig1/fig1_v2.tex}

    \ifvtwofig
    \includegraphics[width=0.49\textwidth]{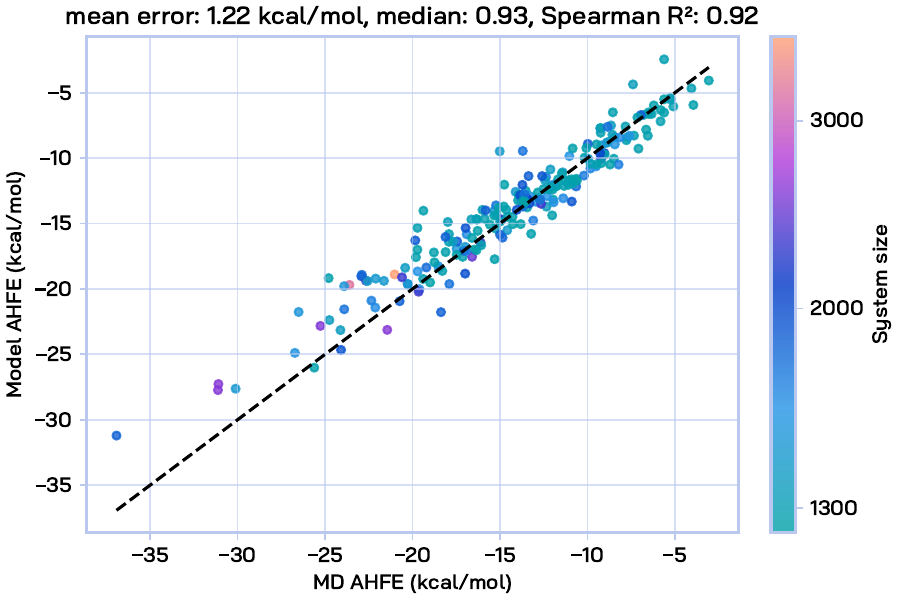}
    \makebox[0pt][l]{\hspace{-0.49\linewidth}\small(b)}
    \includegraphics[width=0.49\textwidth]{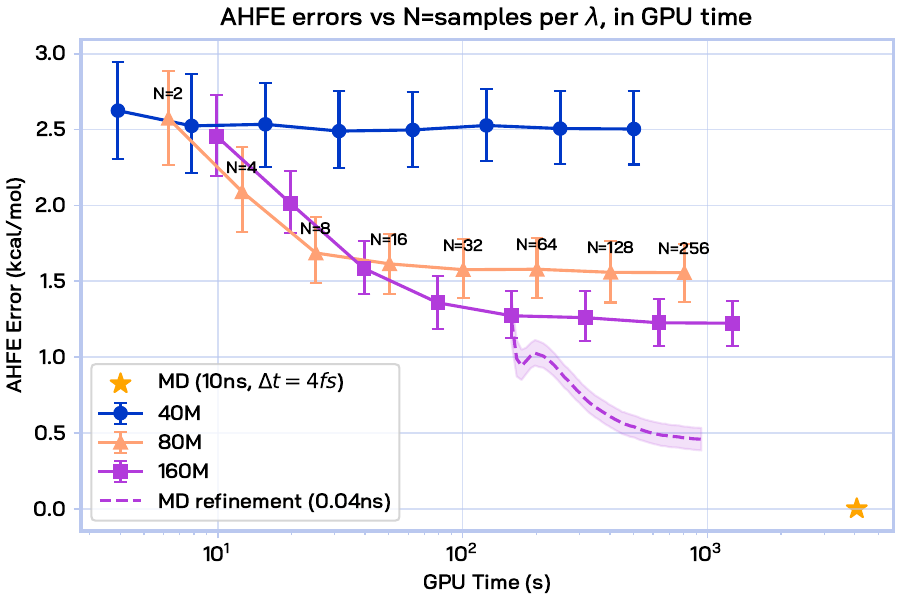}
    \makebox[0pt][l]{\hspace{-0.49\linewidth}\small(c)}
    \else
    \includegraphics[width=0.49\textwidth]{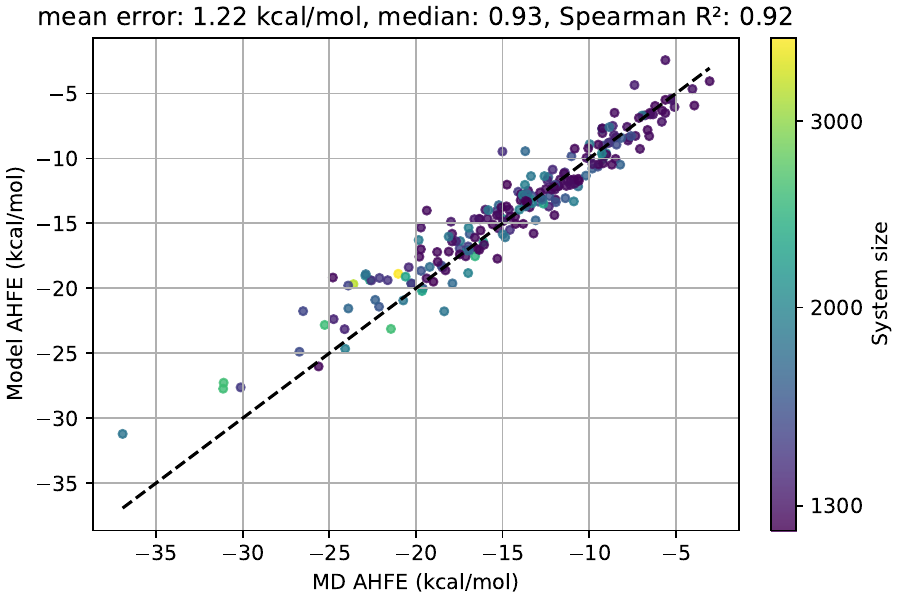}
    \makebox[0pt][l]{\hspace{-0.49\linewidth}\small(b)}
    \includegraphics[width=0.49\textwidth]{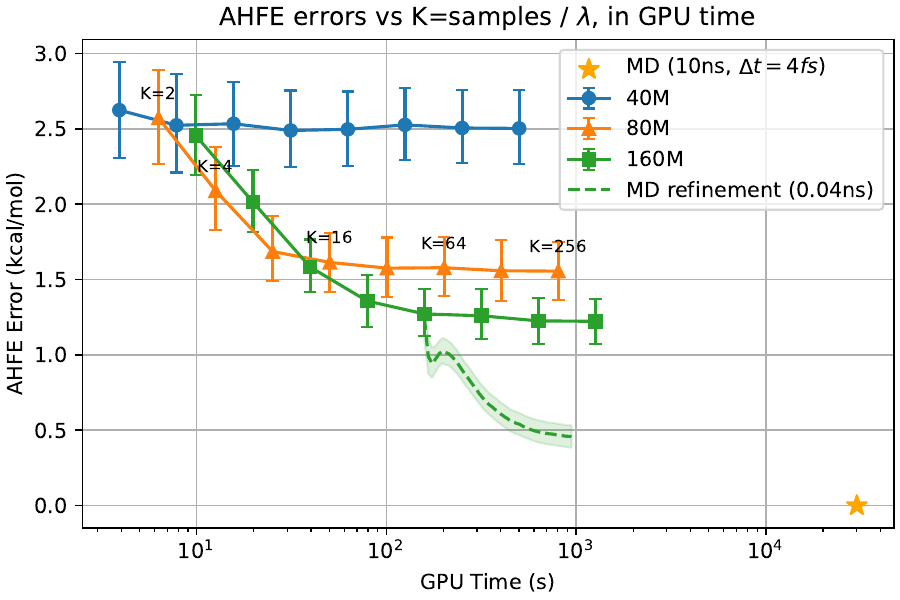}
    \makebox[0pt][l]{\hspace{-0.49\linewidth}\small(c)}
    \fi

    \caption{\textbf{Overview of AquaGen generative modeling framework and results on absolute hydration free energy (AHFE) estimation}. \textbf{(a)} A flow matching model is trained to generate uncorrelated, all-atom, explicit-solvent, PBC-aware configurations $\bar{x}_i^{(\lambda_k)} \sim p_\theta(\cdot \mid \lambda_k), i \in [1, N], k \in [1, K]$ which approximate samples from data distributions $p_{\mathrm{data}}(\cdot \mid \lambda_k)$ along an alchemical pathway governed by $\lambda_k$. Potential energy evaluations $U^{(\lambda_k)}(\bar{x}_i^{(\lambda_j)})$ and MBAR are then used to estimate the absolute hydration free energy (AHFE). \textbf{(b)} Unrefined model samples yield an average and median error of 1.22 kcal/mol and 0.93 kcal/mol respectively on unseen systems, color-coded by number of atoms. \textbf{(c)} Our framework makes use of train- and test-time compute to improve AHFE prediction accuracy. Scaling model size from 40M to 160M parameters improves accuracy, while generating more samples at test-time achieves a similar effect. Operating at all-atom, explicit solvent resolution allows for MD refinement (purple dashed line) starting from generated samples (purple solid line), while remaining 4$\times$ faster than MD.}
    \label{fig:intro_figure}
\end{figure}
\section{Introduction}
Molecular dynamics (MD) is a foundational tool for studying atomistic systems \citep{frenkel2001understanding}. Given a force field describing interatomic interactions, MD simulates the time evolution of a system by numerically integrating the equations of motion at femtosecond resolution. Extracting relevant quantities, such as solvation or binding free energies \citep{gilson2007calculation}, relies on converged sampling of the configurational phase space. In practice, this often requires trajectories spanning nanoseconds to milliseconds \citep{shaw2008anton, lindorff2011fast} , resulting in up to billions of inherently sequential integration steps. Consequently, even modern GPU-accelerated MD simulations can require substantial computational resources while only exploring a limited portion of configurational space. MD nevertheless remains the gold standard for estimating a broad range of structural, dynamical, and thermodynamic properties due to its physical fidelity and generality \citep{shaw2010atomic,wang2015accurate,shirts2007alchemical}.

In this work, we show that using \emph{generative models}, we can produce uncorrelated, approximately Boltzmann-weighted molecular configurations with accuracy on par with MD, and in a much faster and parallelized manner. We introduce \emph{AquaGen}, a generative model which operates at the same resolution as industry-standard bio-molecular MD simulations. We model all atoms, including solvents, and the periodic boundary conditions under which these simulations operate. As a demonstration of the paradigm, we focus on sampling the conformational landscape of small, drug-like molecules solvated in water. The scale of these systems is of the order of $10^3$ atoms. While existing models \citep{abramson2024accurate, passaro2025boltz, lewis2025scalable, kim2024scalable} operate at a similar scale, they either do not account for water, use implicit water, or operate on a lower-dimensional representation (e.g. considering only backbone or heavy atoms). As such, the outputs of these models are either not immediately compatible with high-fidelity energy functions, or require significant post-processing (e.g., energy minimization) before being used in downstream calculations like free energy perturbations (FEP). By contrast, AquaGen is, to our knowledge, the first generative model whose generated configurations are sufficiently geometrically accurate and diverse that they closely reproduce the energy distribution of reference MD simulations under an explicit-solvent, all-atom force field.

Faithfully reproducing the all-atom Boltzmann distribution unlocks the ability to compute pharmaceutically relevant quantities as ensemble averages of potential energy evaluations over generated samples. In this work, we demonstrate the value of this approach in predicting the \emph{absolute hydration free energy (AHFE)}: the Gibbs free energy change associated with transferring a compound from vacuum to water solvent. We train AquaGen to sample conformations of water-solvated, drug-like compounds, conditioned on an alchemical order parameter $\lambda$ which gradually attenuates interactions between the compound and the solvent. We then evaluate potential energies of the generated structures using an OpenFF 2.1.1 force field in conjunction with the TIP3P water model. Average potential energy differences $U^{(\lambda_i)} - U^{(\lambda_j)}$ determine the degree of overlap between samples at different points on the alchemical pathway. These energy differences, computed over many samples and many values of $\lambda$, yield the AHFE via the unbiased multistate Bennett acceptance ratio (MBAR) estimator \citep{shirts2008statistically}. We show that AquaGen is accurate and efficient, producing AHFE estimates at 4-10$\times$ the speed of GPU-based MD, with approximately 1 kcal/mol error. We demonstrate predictable gains from scaling train- and test-time compute, by increasing model capacity and the number of generated samples, respectively.

Since the potential energy is calculated with a physics-based force field, we deem it a \emph{white-box} component of our pipeline. Meanwhile, we consider the generative model itself to be a \emph{black-box} entity, as it is a flexibly parameterized, learnable function approximator. We thus characterize our overall AHFE prediction framework as \emph{gray-box}, in contrast with the growing suite of methods on purely black-box chemical property prediction \citep{chithrananda2020chemberta, ahmad2022chemberta, heid2023chemprop, passaro2025boltz}. Several advantages naturally arise from this gray-box characteristic. For instance, property predictions from AquaGen are \emph{refinable}. Because our model generates physically valid atomistic conformations that are directly compatible with potential energy evaluations, they can serve as starting points for short MD refinement to improve accuracy. Additionally, uncertainty estimates derived from simple bootstrapping over generated samples are well-calibrated with AHFE prediction errors, which may be critical for trust and widespread adoption. If we push this gray-box framework to its logical conclusion, we may be able to scale to more complex atomic systems with $10^4 - 10^5$ atoms and compute quantities such as protein-ligand binding energy \citep{gilson2007calculation} in an inspectable, interpretable, refinable, and reliable manner.

To summarize, the main contributions of this work are the following:
\begin{enumerate}
    \item We introduce \emph{AquaGen}, which is to our knowledge the first all-atom, explicit water molecular generative model of Boltzman-distributed atomistic conformations, scaling up to $4 \times 10^3$ atoms. 
    \item We achieve high energetic and structural accuracy relative to the true Boltzmann distribution of solvated compounds, obtained from over 1 billion frames of alchemical MD trajectories.
    \item By conditioning AquaGen on an alchemical order parameter $\lambda$, we compute absolute hydration free energies (AHFE) of held-out compounds with approximately 1 kcal/mol error, with a 4-10x speedup relative to MD.
    \item We demonstrate that AHFE predictions derived from AquaGen are refinable via short MD simulations to further reduce error to < 0.5 kcal/mol, and that simple, zero-shot uncertainty quantification techniques are well-calibrated with model errors.
\end{enumerate}

\section{Molecular Dynamics Simulation and Free Energy Calculations}

We briefly review the major technical concepts relevant to this work. In \S\ref{sec:md_intro}, we begin by reviewing concepts relevant to MD simulation. In \S\ref{sec:ahfe_intro}, we review how alchemical MD simulations can be used to compute the \emph{absolute hydration free energy} (AHFE) of a compound. The computational expense of these simulations motivates the development of surrogate generative models, which will be introduced in \S\ref{sec:gen_modeling_intro}.

\subsection{Molecular Dynamics Simulation}
\label{sec:md_intro}

MD simulates the time-evolution of a molecular system under a potential energy function. Formally, denote the coordinates of an $N$-atom molecular system as $x \in \mathbb{R}^{N \times 3}$, and the potential energy function as $U(x): \mathbb{R}^{N \times 3}\rightarrow \mathbb{R}$. The dynamics evolve according to Newton's equations of motion, where the force on each atom is given by the negative gradient of the potential energy:
\begin{align}
    m_i \frac{d^2 x_i}{dt^2} = F_i(x) = -\nabla_{x_i} U(x),
\end{align}
for atom $i$ with mass $m_i$. In practice, these equations are numerically integrated using discrete timesteps to generate trajectories $\{x_t\}_{t=0}^T$. In this work, we consider MD of solvated, drug-like compounds obtained from the early stages of internal drug discovery campaigns. To reflect common experimental conditions, we perform constant-temperature, constant-pressure (NPT) MD simulations. In principle under an appropriate choice of thermostat and barostat, NPT MD simulation ergodically samples from the isothermal-isobaric Boltzmann distribution:
\begin{align}
    \label{eq:boltzmann_def}
   p_{\text{NPT}}(\bar{x}) = \frac{\exp(-\beta(U(\bar{x}) + PV(\bar{x})))}{Z},
\end{align}
meaning that a time average of any observable $\hat{O} = \lim_{T \rightarrow \infty} \frac{1}{T} \sum_{t=1}^T O(\bar{x}_t)$ over a sufficiently long simulation converges to an unbiased expectation over the Boltzmann distribution $\mathbb{E}_{\bar{x} \sim p_{\text{NPT}}} O(\bar{x})$. Here,  $\bar{x} = \{x, c\}$, where $c \in \mathbb{R}^{3 \times 3}$ represents the simulation box,  $V: \mathbb{R}^{3 \times 3}\rightarrow \mathbb{R}$ is the instantaneous volume, $\beta = \frac{1}{k_BT}$ is the inverse temperature, $P$ is the pressure, and $Z$ is the partition function.

\subsection{Absolute Hydration Free Energy}
\label{sec:ahfe_intro}
Many problems in drug discovery reduce to estimating free energy differences $\Delta G$, which characterize the extent to which a molecular process, such as solvation or binding, is energetically favorable \citep{cournia2017relative, muegge2023recent}. The Gibbs free energy of a state can be expressed in terms of its partition function as $G = -\beta^{-1} \log Z$, such that the difference in free energy between two states $A$ and $B$ is
\begin{align}
    \Delta G_{A \rightarrow B}
    = G_B - G_A
    = -\beta^{-1} \ln \frac{Z_B}{Z_A}.
\end{align}
A particularly important example is the absolute hydration free energy (AHFE), which measures the free energy change associated with transferring a compound from vacuum into solvent \citep{mobley2014freesolv}. AHFE is a key component of most downstream biomolecular binding calculations, which typically involve transfer of a compound from solution to a binding site. For this reason, AHFE prediction accuracy is often viewed as an approximate upper bound on accuracy for more complex binding free energy prediction tasks. The fact that free energy differences are \emph{state functions}---that is, they depend only on the endpoints and not on the paths connecting them---can be exploited to compute $\Delta G_\text{AHFE}$ as
\begin{align}
    \label{eq:solv_minus_vacuum}
    \Delta G_\text{AHFE}
    =
    \Delta G_\text{vacuum}
    -
    \Delta G_\text{solvated},
\end{align}
where $\Delta G_\text{vacuum}$ and $\Delta G_\text{solvated}$ are the free energy changes associated with annihilating intermolecular interactions in vacuum and solvent respectively. The first term is typically cheap to compute, as it only involves a single drug-like molecule in vacuum. The second term is the computational bottleneck due to the presence of solvent and the resulting high-dimensional configurational space. Focusing on the solvated component, the free energy difference can formally be written in terms of a ratio of partition functions:
\begin{align}
    \Delta G_\text{solvated}
    =
    -\beta^{-1}
    \log
    \frac{
        Z_{\text{solvated, interacting}}
    }{
        Z_{\text{solvated, non-interacting}}
    }.
\end{align}
These partition functions are intractable, as they require integration over all configurational degrees of freedom of the compound and solvent. Instead, one can leverage MD to sample from the corresponding Boltzmann distributions. The Zwanzig formula
\begin{align}
\label{eq:zwanzig}
    \Delta G_{\text{solvated}}
    &=
    -\beta^{-1}
    \log
    \mathbb{E}_{\bar{x} \sim p_{\text{solvated, non-interacting}}}
    \left[
    \exp\left(
    -\beta\left(
    U_\text{solvated, interacting}(\bar{x})
    -
    U_\text{solvated, non-interacting}(\bar{x})
    \right)
    \right)
    \right]
\end{align}
gives an unbiased estimator of the solvation free energy difference, with expectations approximated via samples from MD \citep{zwanzig1954high}. However, this estimator typically has impractically high variance due to the low overlap between the distributions
$p_{\text{solvated, non-interacting}}$
and
$p_{\text{solvated, interacting}}$.
Converged free energy estimates would require potentially unfeasibly long MD simulations to adequately sample the extremely rare regions of overlap between these distributions. To alleviate this issue, one can introduce a sequence of overlapping intermediate alchemical distributions indexed by $\{\lambda_k\}_{k=1}^K$ (each with potential energy $U^{(\lambda_k)}$) that gradually transition from a fully interacting to a fully non-interacting system. The multistate Bennett acceptance ratio (MBAR) estimator~\citep{shirts2008statistically} can be used to pool samples from all $\lambda$-states and compute an unbiased, minimum-variance estimate of the relative free energies. Thus, estimating $\Delta G_\text{solvated}$ reduces to running multiple $\lambda$-dependent MD simulations and computing MBAR on the samples.

\paragraph{Computational Cost.} MD simulation offers a principled mechanism by which to sample from the Boltzmann distribution, and alchemical simulations can be used to accurately compute relevant free energy differences such as AHFE. However, due to presence of energy barriers separating relevant regions of the potential energy surface, MD requires long time horizons for relevant expectation values to converge. Considering the small time discretization necessary for numerical stability, along with the iterative and non-parallelizable nature of the algorithm, this makes alchemical MD simulation expensive in wall clock time--on the order of 1 GPU-hour per $\lambda$ for a 1000-atom system. This is further exacerbated by the fact that the simulation length necessary for convergence can be a strong function of the system of interest, meaning that in practice we may need to run long simulations for all systems to ensure convergence. 
\section{AquaGen: All-Atom, Explicit-Water Generative Modeling}
\label{sec:gen_modeling_intro}
To alleviate the challenges posed by all-atom, explicit-water MD simulations, we turn to \emph{generative modeling}. 

\subsection{Generative Modeling Framework}

The central idea is to train a model which can generate Boltzmann-distributed configurations that are energetically and structurally indistinguishable from MD samples, without slow, iterative traversal of configurations. Formally, given a data distribution $p_{\text{data}} \approx p_{\text{NPT}}$ obtained from classical MD simulation, we wish to parameterize a generative model $f_\theta$ which maps a sample $z \sim p_0$ from a tractable prior distribution, $p_0$, to a sample $x = f_\theta(z) \sim p_\theta \approx p_{\text{data}}$ matching an i.i.d draw from $p_\text{data}$. The generative modeling framework straightforwardly extends to sampling from alchemical Boltzmann distributions $\{p_{\text{data}}(\cdot \mid \lambda)\}_{\lambda \in [0,1]}$ simply by conditioning the generative model on the alchemical parameter $\lambda$. Concretely, the $\lambda$-conditional model $f_\theta(\cdot | \lambda)$ learns to generate configurations $\bar{x} = \{x, c\}$ distributed according to $p_\text{data}(\cdot| \lambda)$, where the $[0,1]$ interval is discretized into $K$ lambda values  $\{\lambda_k\}_{k=1}^K$. After training, we draw $N$ samples $\{\bar{x}_i^{(\lambda_k)}\}_{i=1}^N \sim p_\theta(\cdot \mid \lambda_k)$ for $k \in [1, K]$ from the generative model and follow the rest of the AHFE estimation procedure described in \S\ref{sec:ahfe_intro} as if we had obtained samples from long, alchemical MD simulations:  evaluate reduced energies $\{u^{(\lambda_k)}(\bar{x}_i) \}_{i=1}^N = \{\beta U^{(\lambda_k)}(\bar{x}_i) \}_{i=1}^N$, and pass these into MBAR to obtain an estimate of $\Delta G_{\text{solvated}}$. For the vacuum contribution $\Delta G_{\text{vacuum}}$, we rely on classical MD estimates, as sampling in vacuum is computationally inexpensive relative to the solvated setting. The final AHFE is computed from ~\eqref{eq:solv_minus_vacuum}.

\paragraph{Advantages of Generative AHFE Estimation.} Unlike regression models which make \emph{black-box} predictions, the generative route is \emph{gray-box}; the AHFE prediction is derived via actual evaluations of potential energy functions $U^{(\lambda_k)}$ on the generated samples, which have interpretable physical meaning. Due to the use of MBAR, reasonable estimates of AHFE can only stem from reasonable overlaps in the energy distributions of adjacent $\lambda$ values, and by extension from physical Boltzmann-weighted configurations.  As we will show in \S\ref{sec:ahfe_results}, computing AHFE via generative model samples also introduces a natural axis of test-time compute scaling to increase prediction accuracy: the number of generated configurations per $\lambda$ value. The paradigm also enables \emph{refinability}, meaning it is possible to further increase accuracy by initiating very short MD simulations in parallel from the generated samples, and zero-shot \textit{uncertainty quantification} via bootstrapped confidence intervals.

\subsection{Generative Modeling Details}
\label{sec:generative_modeling_details}
We use flow matching \citep{lipman2023flow}, a class of generative models which learns a transport map between a tractable prior distribution $p_0$ and a target data distribution $p_{\text{data}}(\cdot \mid \lambda)$. Specifically, flow matching parameterizes a time-dependent ``velocity" field $v_\theta(\bar{x}_\tau, \tau | \lambda)$ which defines a deterministic flow via the ordinary differential equation
\begin{align}
    \label{eq:flow_ode}
    \frac{d \bar{x}_\tau}{d\tau}
    =
    v_\theta(\bar{x}_\tau, \tau | \lambda),
    \quad
    \bar{x}_0 \sim p_0.
\end{align}
We emphasize that the flow matching velocity field is not a physical velocity, but rather is a rate of change of probability along the conditional path. We also use $\tau$ to denote the flow matching time in order to differentiate it from the notion of physical time $t$ in MD simulations. The model is trained to match a target conditional velocity field $v^\star(\bar{x}_\tau, \tau | \lambda)$ induced by a prescribed conditional probability path between $p_0$ and $p_{\text{data}}(\cdot \mid \lambda)$, by minimizing
\begin{align}
    \mathcal{L}(\theta)
    =
    \mathbb{E}_{\tau, \lambda, \bar{x}_\tau}
    \left[
    \|
    v_\theta(\bar{x}_\tau, \tau | \lambda)
    -
    v^\star(\bar{x}_\tau, \tau | \lambda)
    \|^2
    \right].
\end{align}

We perform flow matching jointly over Euclidean coordinates $x$ and the simulation cell $c$. We use a Gaussian prior $p_0(x) = \mathcal{N}(0,2I_{3N})$ for the Euclidean coordinates (Figure \ref{fig:intro_figure}a). For the cell, we adopt a cubic Gaussian prior; that is, we sample a box length from $\ell \sim \mathcal{N}(\mu_c, \sigma^2_c)$ and construct a cell from mutually orthogonal lattice vectors of length $\ell$. We adopt a simple, linear conditional probability path for training:
\begin{align}
    \bar{x}_\tau
    =
    \tau \bar{x}_1
    +
    (1-\tau)\bar{x}_0,
\end{align}
where $\bar{x}_1 \sim p_{\text{data}}(\cdot \mid \lambda)$ and $\bar{x}_0 \sim p_0$, and addition and multiplication are performed independently for the Euclidean coordinates and simulation box. The target conditional velocity for this choice of probability path is given by
\begin{align}
    v^\star(\bar{x}_\tau, \tau | \lambda)
    =
    \bar{x}_1 - \bar{x}_0.
\end{align}

At inference time, we draw a prior sample from $p_0(\bar{x})$ and solve the ODE in \eqref{eq:flow_ode} with Euler integration over $\tau \in [0,1]$ to produce an approximate sample from $p_\text{data}(\bar{x} \mid \lambda)$. We find that the learned vector field has higher curvature near the prior, so we use a finer discretization near $\tau \approx 0$ (see \S\ref{app:velocity_curvature} for more details). 

\paragraph{Model architecture.} We parameterize the velocity field $v_\theta$ with a standard graph neural network (GNN).
The model's input coordinates are encoded as pairwise displacement vectors between nodes in the graph. While we account for all atoms in the input featurization and final velocity prediction head, in the core message-passing block we represent water O-H-H triplets with a single node to reduce overhead for large systems dominated by solvent molecules. We model periodic boundary conditions induced by the simulation box as additional virtual nodes. For more details on the model architecture, see \S\ref{app:arch}.

\subsection{Molecular Dynamics Training Data}
\label{sec:data}
We use a dataset comprising several thousand drug-like compounds derived from internal medicinal chemistry lead optimization. For each compound, we ran alchemical MD simulations with 20 discrete $\lambda$ values with Hamiltonian replica exchange (HREX). The alchemical pathway uses a partial annihilation scheme in which electrostatic interactions are fully annihilated prior to Lennard-Jones interactions being decoupled \citep{alibay2026openfe}. We extracted uncorrelated samples from production trajectories of each $\lambda$ value and each compound, yielding over 1 billion training frames. Hydration free energies were computed using MBAR as implemented in PyMBAR \citep{shirts2008statistically}. See \S\ref{app:appendix_data} for more details about the training data.

\section{Results}

We first present the highest-level, end-to-end AHFE estimation results from Figure \ref{fig:intro_figure} in \S \ref{sec:ahfe_results} . We then zoom in and analyze the energetic and structural accuracy of the generated samples at the fully-interacting endpoint (\S \ref{sec:fully_interacting_energetic_structural_accuracy}) and the intermediate alchemical distributions (\S \ref{sec:alchemical_energy_structure_accuracy}). In \S \ref{sec:ahfe_additional_results}, we discuss more nuanced aspects of AHFE estimation with AquaGen, such as uncertainty estimation and error cancellation. Finally, we perform various ablations in \S \ref{sec:ablations}.
\subsection{Absolute Hydration Free Energy Estimation}
\label{sec:ahfe_results}
We use the generated samples at each $\lambda$ value along the alchemical pathway to compute an unbiased estimate of $\Delta G_{\text{solvated}}$ via the MBAR estimator. This is combined with $\Delta G_{\text{vacuum}}$ from classical MD to produce a final AHFE prediction (\eqref{eq:solv_minus_vacuum}). For a given compound, the AHFE absolute error (AE) is obtained as the absolute difference between the predicted AHFE, computed via MBAR on the generated samples, and the true AHFE, computed via MBAR on reference MD samples. Formally, given samples across $K$ alchemical states, MBAR returns a matrix $\Delta \hat{G} \in \mathbb{R}^{K \times K}$ representing the estimated pairwise free energy differences between states. We compute $\Delta \hat{G}^\text{model}$ and $\Delta \hat{G}^\text{MD}$ from the generative model and MD samples respectively. The AHFE AE is computed as
\begin{align}
\label{eq:ahfe_ae}
\text{AHFE AE}
&= |\Delta \hat{G}^\text{model}_{1, K} - \Delta \hat{G}^\text{MD}_{1, K}|
\end{align}

\paragraph{Prediction accuracy.} Figure \ref{fig:intro_figure}b shows AHFE values resulting from MBAR on samples from AquaGen (y-axis) and ground truth alchemical MD simulations (x-axis) for 218 held-out compounds not seen during training. We achieve a median and mean AHFE AE of 0.93 kcal/mol and 1.22 kcal/mol relative to MD, respectively. More water-soluble compounds (i.e., those with a more negative AHFE) generally have higher positive errors. Given \eqref{eq:solv_minus_vacuum}, this translates to an \textit{under-estimation} of the change in solvation free energy $\Delta G_{\text{solvated}}$. For highly soluble compounds with energetically favorable electrostatic interactions between the compound and solvent, we hypothesize that due to suboptimal conditioning on the alchemical parameter $\lambda$, the model learns an ``averaged" distribution over $\lambda_k, k \in [1, 5]$ which yields larger overlaps and a smaller $\Delta G_{\text{solvated}}$ (we substantiate this hypothesis further in \S \ref{sec:ahfe_additional_results} when we discuss \emph{error cancellation}).

\begin{figure}
    \centering
    \ifvtwofig
    \includegraphics[width=0.3\linewidth]{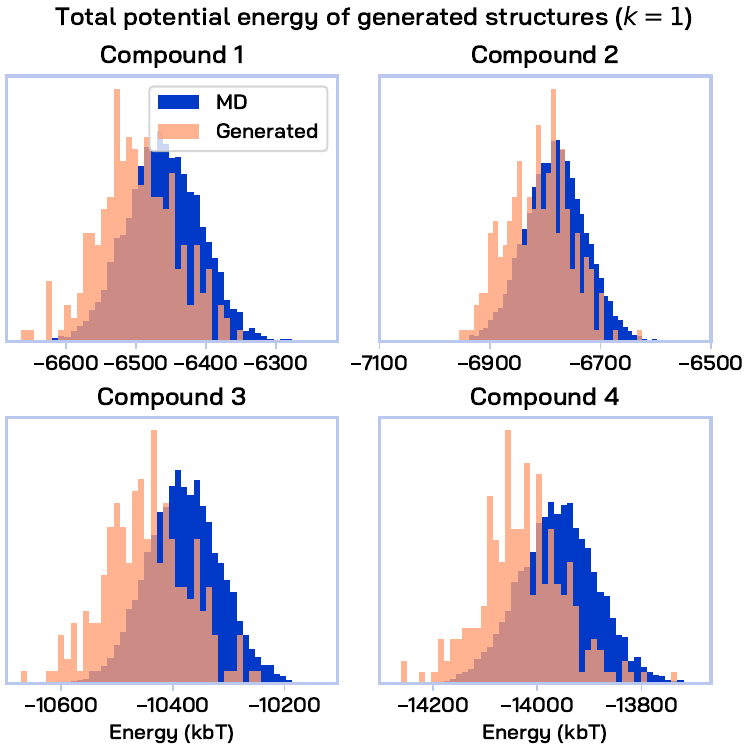}
    \makebox[0pt][l]{\hspace{-0.32\linewidth}\small(a)}
    \includegraphics[width=0.3\linewidth]{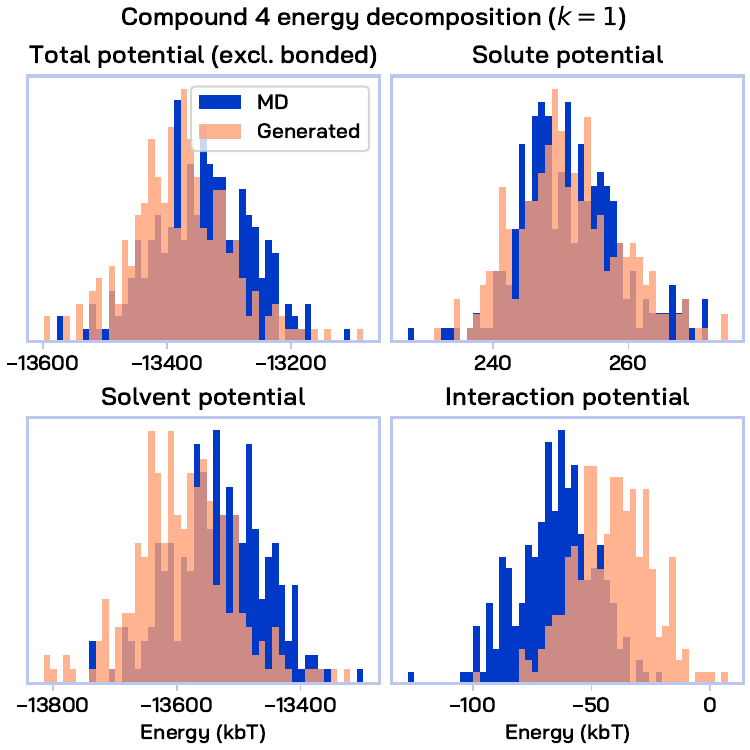}
    \makebox[0pt][l]{\hspace{-0.32\linewidth}\small(b)}
    \includegraphics[width=0.3\linewidth]{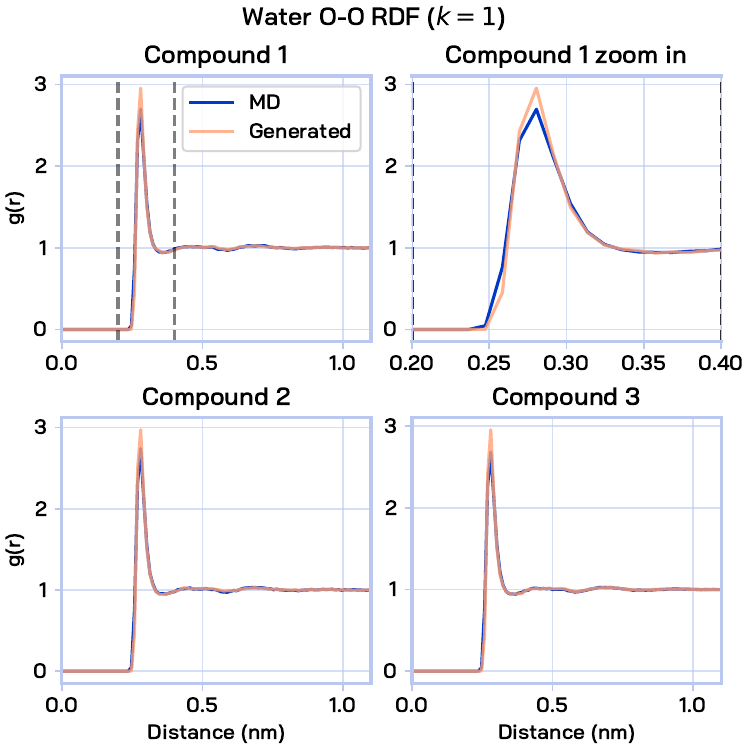}
    \makebox[0pt][l]{\hspace{-0.32\linewidth}\small(c)}
    \else
    \includegraphics[width=0.3\linewidth]{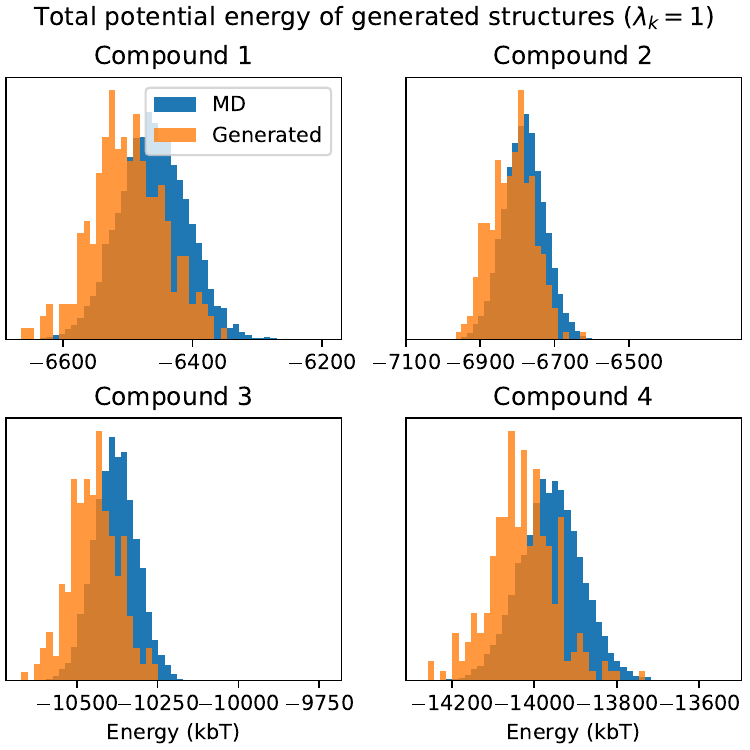}
    \makebox[0pt][l]{\hspace{-0.32\linewidth}\small(a)}
    \includegraphics[width=0.3\linewidth]{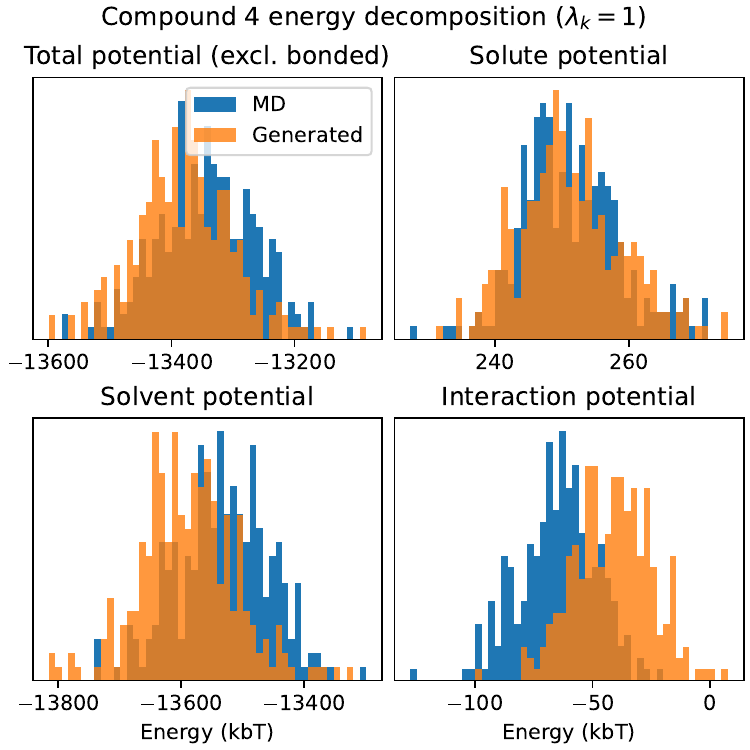}
    \makebox[0pt][l]{\hspace{-0.32\linewidth}\small(b)}
    \includegraphics[width=0.3\linewidth]{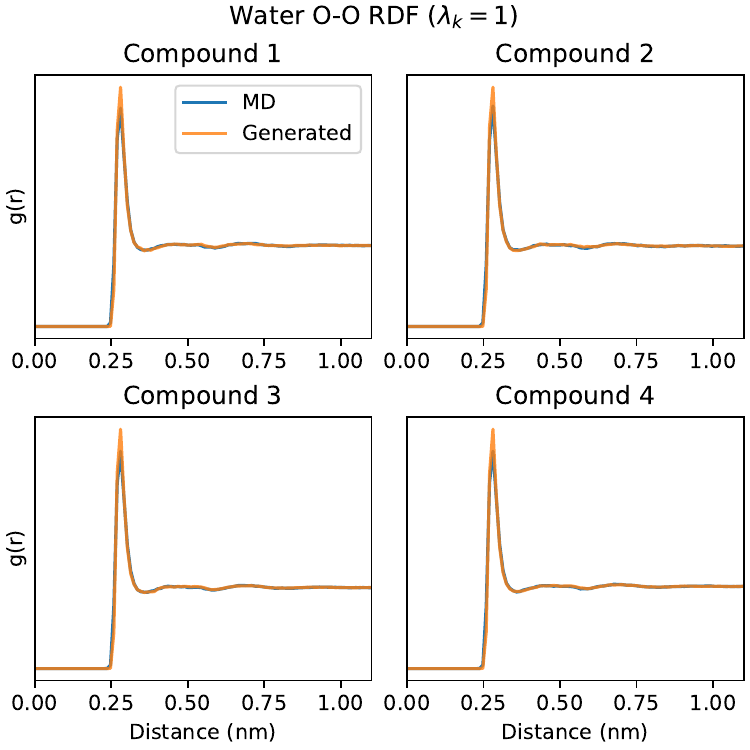}
    \makebox[0pt][l]{\hspace{-0.32\linewidth}\small(c)}
    \fi
    \caption{\textbf{Structural and energetic accuracy of samples generated by AquaGen at the fully interacting endpoint ($\lambda_k, k = 1$)} \textbf{(a)} True vs generated potential energy for 4 randomly chosen compounds. \textbf{(b)} Various decompositions of compound 4 energy. In clockwise order, starting from top left: total potential energy excluding solvent bonded terms (since the reference simulations are performed with rigid water), solute-only potential energy, solute-solvent interaction potential energy, solvent-only potential energy. All generated distributions are strongly overlapping with reference distributions, with slight overestimation of the solute-solvent interaction energy. \textbf{(c)} True and generated water O-O radial distribution functions (RDFs). The generated RDF is slightly overly ordered relative to the MD reference.}
    \label{fig:interacting_dist_accuracy}
\end{figure}

\paragraph{Train/test-time scaling and refinability.} Our approach naturally admits train- and test-time compute scaling to improve AHFE estimation accuracy. Train-time compute is scaled by increasing model capacity, while test-time compute is scaled by increasing the number of generated samples from 2 to 256 at each $\lambda_k$. Figure \ref{fig:intro_figure}c shows mean AHFE AE as a function of test-time GPU compute per $\lambda$ value, per compound, with curves for varying model sizes. For any model size, generating more samples leads to improved results. The performance of smaller models plateaus more quickly, while larger models continue to improve up to 256 samples. Scaling model size from 40M to 160M parameters leads to improvements in AHFE at larger test-time compute budgets, while all models are relatively similar at smaller compute budgets. This suggests that larger models produce more diverse samples. By initiating very short (40 ps) MD refinements from the generated samples of the 160M parameter model, we achieve an AHFE error of 0.5 kcal/mol relative to MD, still with about 4x less GPU time (we use the same settings for the MD refinement as the data generation, see \S \ref{app:appendix_data}). Note that generative model sampling is trivially parallelizable across $\lambda$-windows, while MD incurs communication overhead for parallelization across $\lambda$ due to the replica exchange, and most importantly is not parallelizable across time. Thus, given more inference hardware resources, in the future we can expect greater speedups in \emph{wall clock time} from generative modeling relative to MD. 
\begin{figure}[!ht]
    \centering
    \ifvtwofig
    \includegraphics[width=0.9\linewidth]{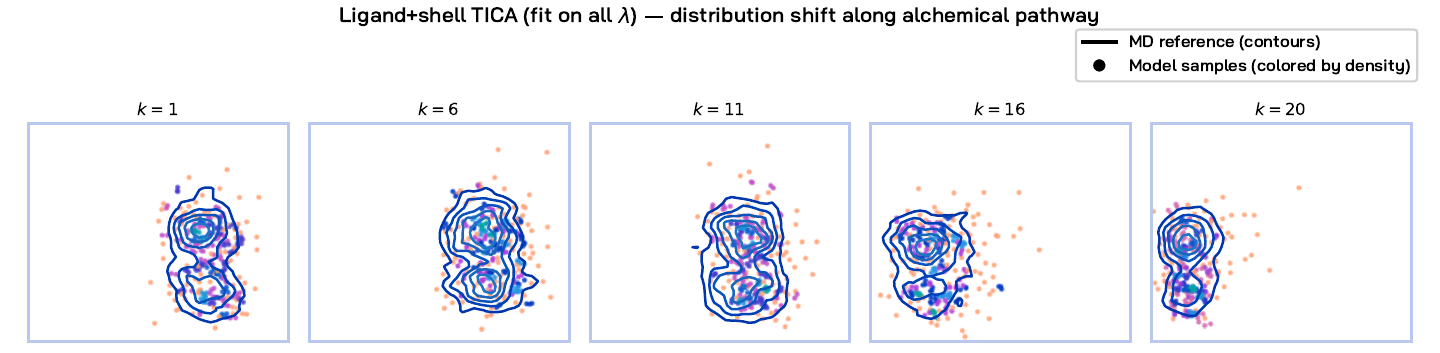} 
    \makebox[0pt][l]{\hspace{-0.915\linewidth}\small(a)}\\
    \includegraphics[width=0.30\linewidth]{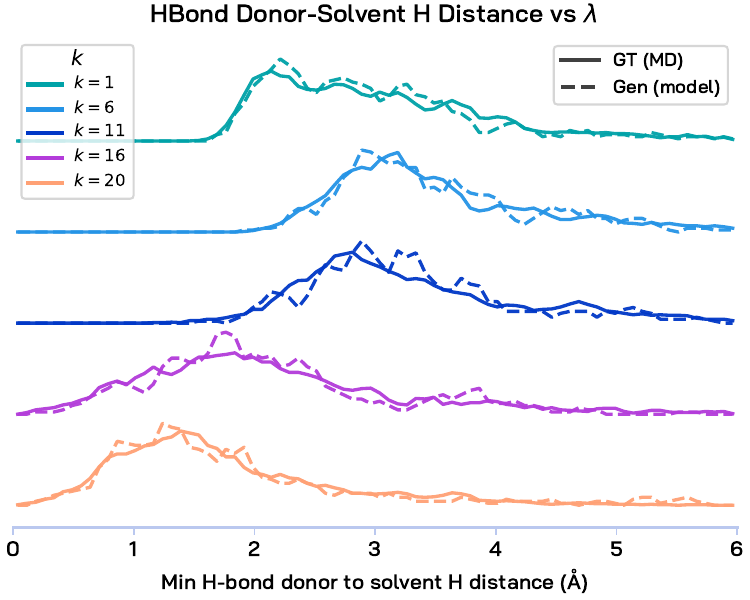}
    \makebox[0pt][l]{\hspace{-0.30\linewidth}\small(b)}
    \includegraphics[width=0.30\linewidth]{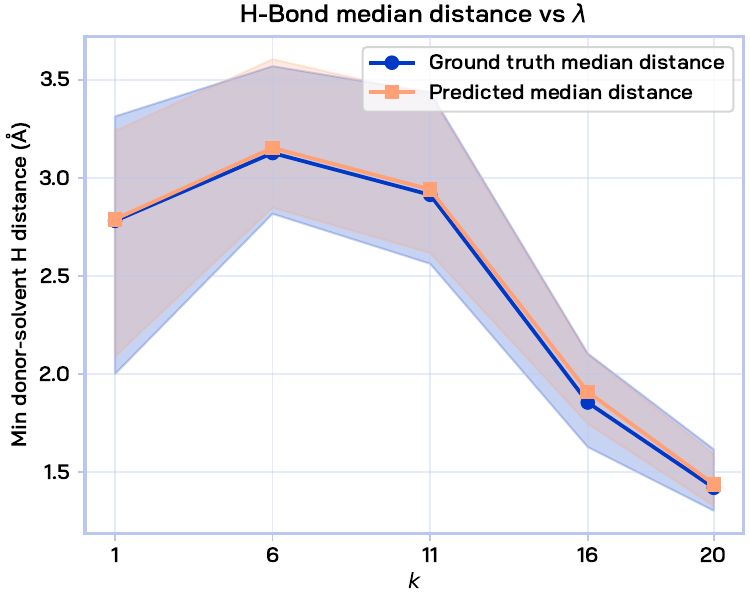}
    \makebox[0pt][l]{\hspace{-0.30\linewidth}\small(c)}
    \includegraphics[width=0.30\linewidth]{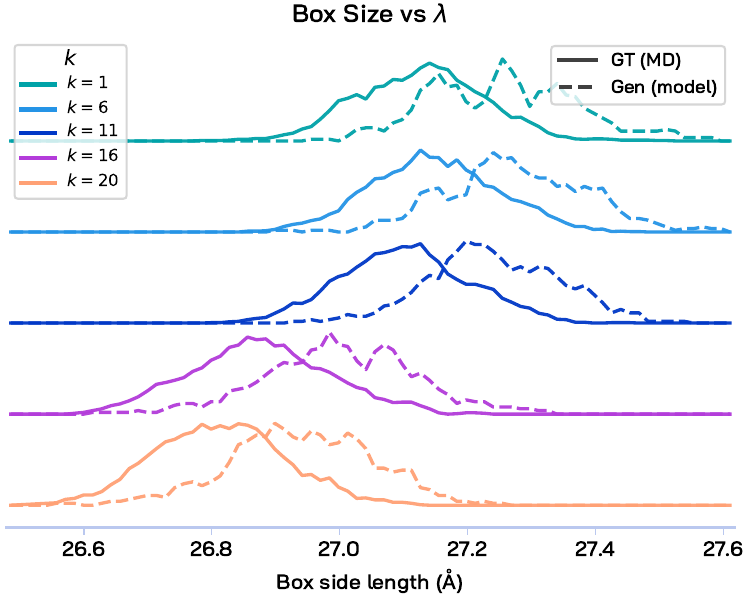}
    \makebox[0pt][l]{\hspace{-0.30\linewidth}\small(d)}
    \else
    \includegraphics[width=0.9\linewidth]{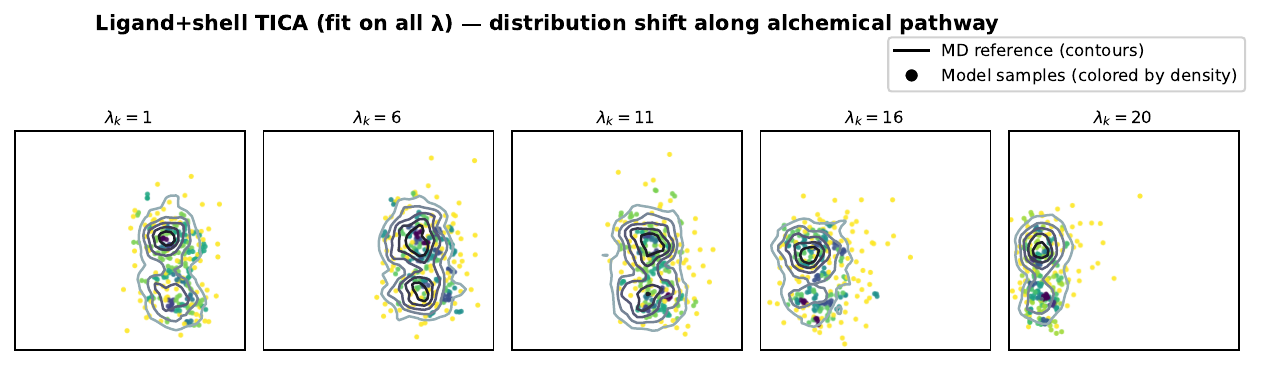} 
    \makebox[0pt][l]{\hspace{-0.9\linewidth}\small(a)}\\
    \includegraphics[width=0.30\linewidth]{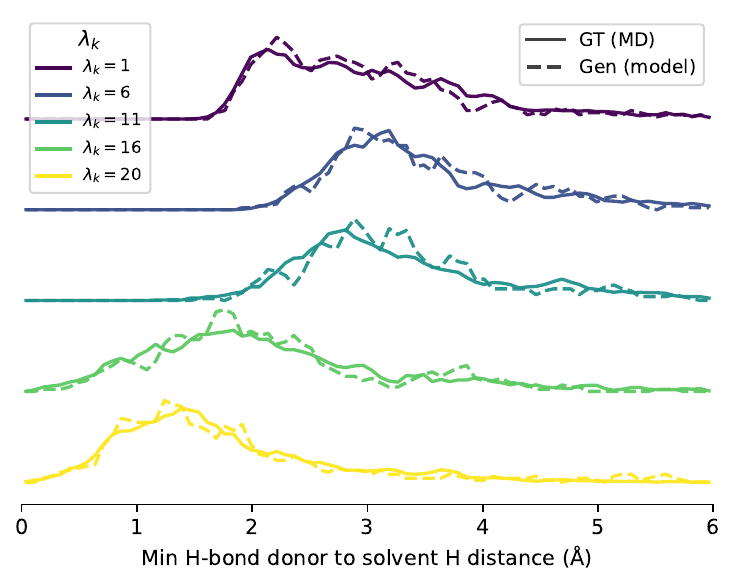}
    \makebox[0pt][l]{\hspace{-0.30\linewidth}\small(b)}
    \includegraphics[width=0.30\linewidth]{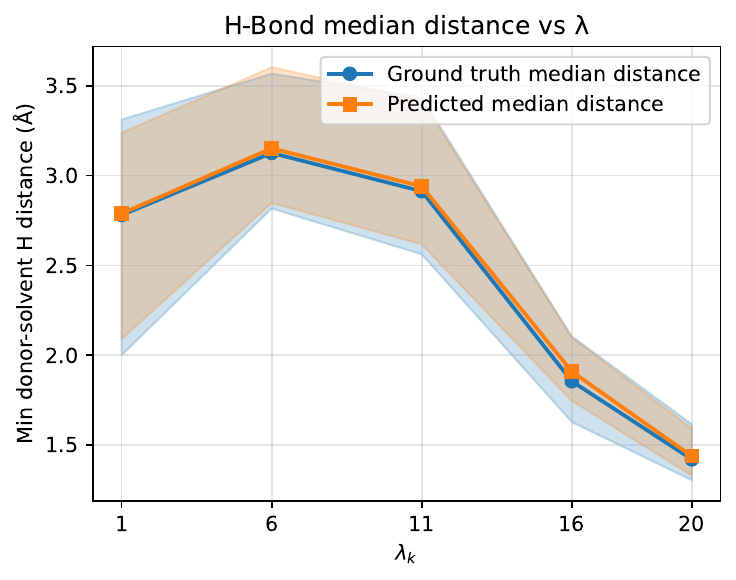}
    \makebox[0pt][l]{\hspace{-0.30\linewidth}\small(c)}
    \includegraphics[width=0.30\linewidth]{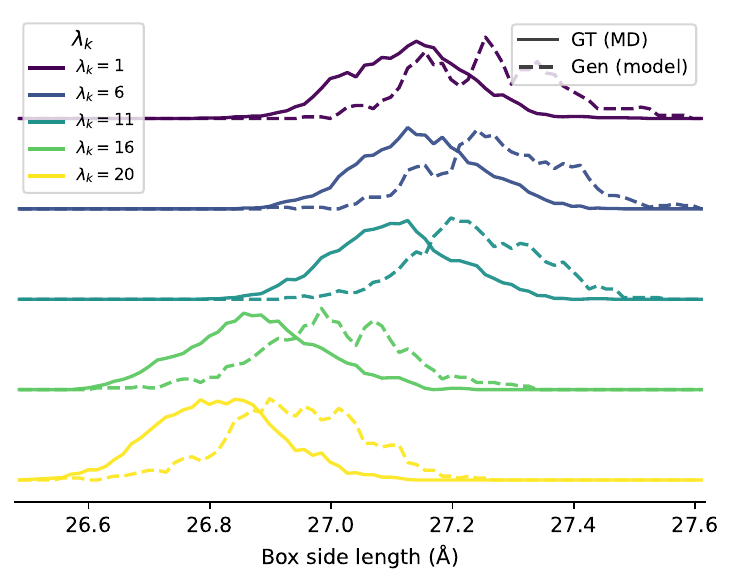}
    \makebox[0pt][l]{\hspace{-0.30\linewidth}\small(d)}
    \fi
    \caption{\textbf{Structural and energetic accuracy of all-atom, explicit water samples generated by AquaGen along the alchemical pathway ($\lambda_k, k \in [1, 20]$).} \textbf{(a)} Time-lagged independent component analysis (tICA) plots of samples from reference MD simulations (contours) and flow matching model (points - darker indicates higher density) across $\lambda_k, k \in \{1, 6, 11, 16, 20\}$. The model samples move across the first tICA component along with the reference contours, indicating sampling of the correct regions of conformational space. \textbf{(b)} Distribution of minimum distance between solvent hydrogen atom and electronegative atom (N, O, F) on the compound, at different lambdas, for a randomly selected compound. The generated samples closely match the non-monotonic trend along the alchemical pathway. \textbf{(c)} The mean across compounds of the minimum distance between solvent hydrogen atom and electronegative atom (N, O, F) on the compound, as a function of $\lambda_k$. The model samples capture the trend closely. \textbf{(d)} Distribution of simulation box lengths of generated and reference samples. While AquaGen captures the correct trend as a function of $\lambda$, it tends to overestimate the box size.}
    \label{fig:alch_dist_accuracy}
\end{figure}

\subsection{Energetic and Structural Accuracy of Fully-Interacting Endpoint}
\label{sec:fully_interacting_energetic_structural_accuracy}

Having assessed the end-to-end AHFE estimation capabilities of AquaGen, we now zoom in and assess our ability to produce samples which closely align with the physical Boltzmann distribution produced by MD. To facilitate this understanding, in this section, we report results from a model trained only on frames from the fully interacting distribution (\eqref{eq:boltzmann_def}). Results are shown in Figure \ref{fig:interacting_dist_accuracy}. The potential energy distribution of generated samples is highly overlapping with the reference distribution produced by MD (Figure \ref{fig:interacting_dist_accuracy}a). No refinement or post-processing (e.g., energy minimization) was performed on the sampled frames. More granular energy decompositions (\ref{fig:interacting_dist_accuracy}b) reveal that the model slightly overestimates the solute-solvent interaction energy. While the ground truth MD data uses a rigid water assumption, AquaGen has no such constraint on the positions of generated water atoms.  Thus, we have excluded bonded solvent energy contributions in Figure \ref{fig:interacting_dist_accuracy}b on the top left (see \S\ref{app:appendix_results} for more detailed analysis). We also find that the bulk-water radial distribution functions (RDFs) produced by the model are quite accurate, though slightly overly ordered (Figure \ref{fig:interacting_dist_accuracy}
c). 

\subsection{Energetic and Structural Accuracy of Alchemical Annihilation}
\label{sec:alchemical_energy_structure_accuracy}
We next assess the extent to which AquaGen captures the essential physics of the solute-solvent interaction annihilation along the alchemical pathway $\{p_{\text{NPT}}(\bar{x} \mid \lambda_k)\}_{k \in [1, 20]}$. Results in this section are reported on a model trained on all $\lambda_k, k \in [1, 20]$ (the same model which was used to produce the end-to-end AHFE results in Figure \ref{fig:intro_figure}). Figure \ref{fig:alch_dist_accuracy}a shows a time-lagged Independent Component Analysis (tICA) plot of the reference MD and generated samples across $\lambda$ values, showing that the model samples the correct the regions of conformational space (see \S\ref{app:appendix_results} for tICA plots of more compounds and details on the tICA featurization). We also analyze the distribution of distances between an electronegative atom (N, O, or F) on the compound and the closest hydrogen atom on a solvent water molecule, constituting a potential hydrogen-bond donor-acceptor pair. We expect the mean of this distribution to be a non-monotonic function of the alchemical parameter $\lambda$. In the electrostatic regime ($\lambda_k, k \in [1, 5]$), the annihilation of attractive electrostatic interactions between the compound and solvent should lead to an increase in the mean hydrogen-bond distance, while in the van der Waals regime ($\lambda_k, k \in [6, 20]$), the annihilation of repulsive, steric interactions should lead to a decrease in the mean hydrogen-bond distance. AquaGen captures this trend well, both for a single randomly chosen compound (Figure \ref{fig:alch_dist_accuracy}b) and across many compounds (Figure \ref{fig:alch_dist_accuracy}c). A salient and consistent error of AquaGen is the tendency to overestimate the simulation box length (Figure \ref{fig:alch_dist_accuracy}d). We observe a consistent overestimation of about 0.1~\AA, which we attribute to our choice of prior (see \S \ref{sec:ablations} for details).

\subsection{Additional Analysis of AHFE Predictions}
\label{sec:ahfe_additional_results}
We now discuss more nuanced aspects of estimating AHFE with AquaGen; notably, estimation of uncertainities, generalization away from the training set, error cancellation across the alchemical pathway, and a comparison to black-box prediction methods.

\begin{figure}[t]
    \centering
    \ifvtwofig
    \includegraphics[width=0.32\linewidth]{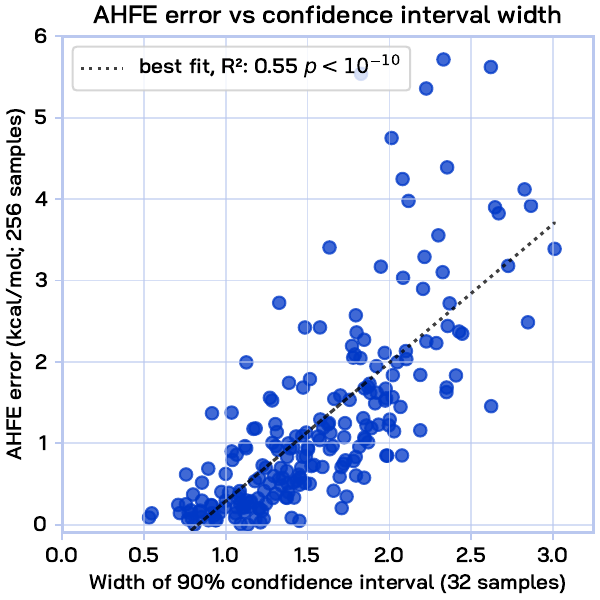}
    \makebox[0pt][l]{\hspace{-0.32\linewidth}\small}
    \includegraphics[width=0.32\linewidth]{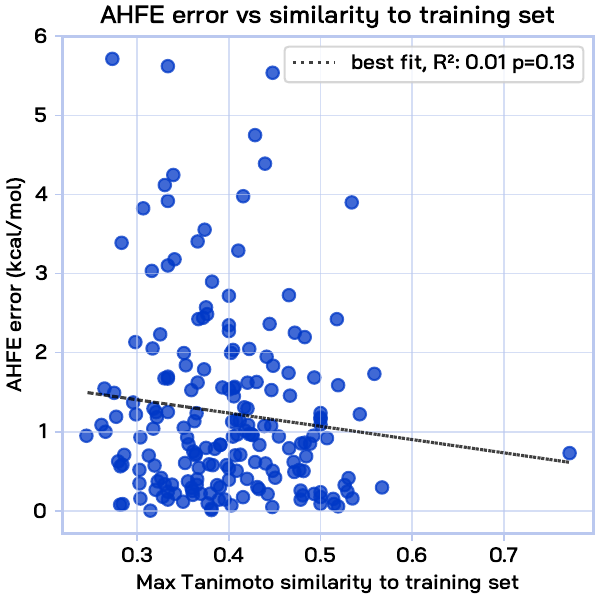}
    \makebox[0pt][l]{\hspace{-0.32\linewidth}\small(b)}
    \includegraphics[width=0.32\linewidth]{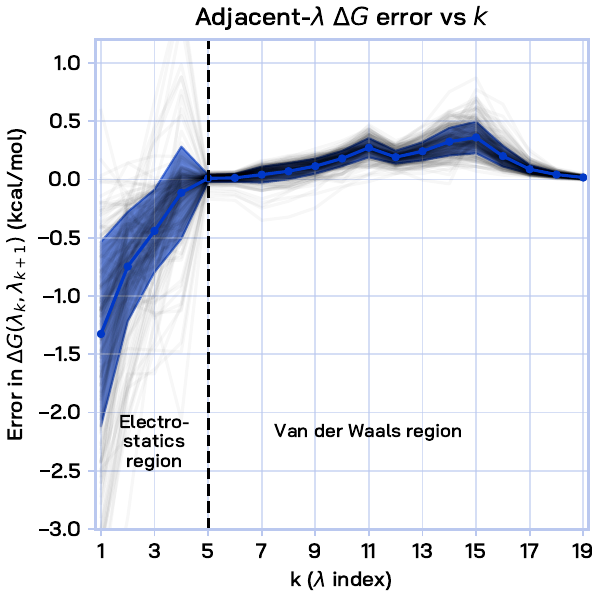}
    \makebox[0pt][l]{\hspace{-0.32\linewidth}\small(c)}
    \else
    \includegraphics[width=0.32\linewidth]{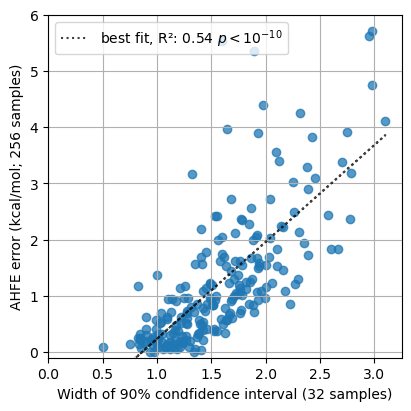}
    \makebox[0pt][l]{\hspace{-0.32\linewidth}\small(a)}
    \includegraphics[width=0.32\linewidth]{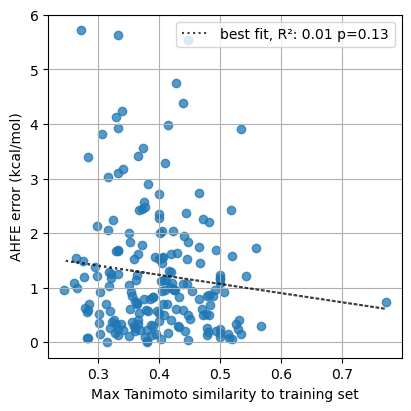}
    \makebox[0pt][l]{\hspace{-0.32\linewidth}\small(b)}
    \includegraphics[width=0.32\linewidth]{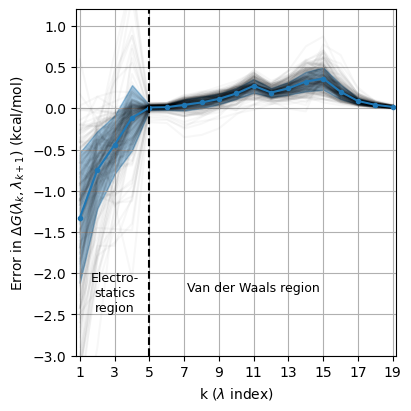}
    \makebox[0pt][l]{\hspace{-0.32\linewidth}\small(c)}
    \fi
    \caption{\textbf{Analysis of uncertainty estimation, generalization across compounds, and error cancellation when using AquaGen for AHFE prediction.}
    \textbf{(a)} The width of bootstrapped 90\% confidence intervals, computed from random 32-sample subsets of 128 generated configurations, correlates strongly with the AHFE prediction error, indicating that AquaGen provides well-calibrated uncertainty estimates. \textbf{(b)} AHFE prediction error as a function of the maximum Tanimoto similarity between each test compound and any compound in the training set. The weak correlation suggests that AquaGen learns transferable representations as opposed to memorizing training compounds.
    \textbf{(c)} Errors in the first off-diagonal of the MBAR $\Delta \hat{G}$ matrix estimated from the reference MD and generated samples. The AHFE Absolute Error (AE) is the cumulative sum (i.e., area under the curve) of this plot. Since errors have opposite signs in different regions of the alchemical pathway, \emph{error cancellation} affects the final AHFE AE. The Cumulative Absolute Error (CAE) metric is the cumulative sum of the \emph{absolute value} of this plot, and is thus robust to error cancellation.
    }
    \label{fig:ahfe_fig}
\end{figure}
\paragraph{Calibrated uncertainties.} As another demonstration of the utility of our gray-box framework for AHFE prediction, we can extract uncertainty estimates from AquaGen at test-time by computing the width of boostrapped confidence intervals on 32-sample subsets of the generated samples. This uncertainty estimate is well-calibrated, correlating strongly with the actual AHFE error (Figure \ref{fig:ahfe_fig}a). This procedure can be thought of as analogous to estimating uncertainties by running multiple MD replicas, but with considerably lower computational cost. Such estimates can be used to guide downstream decisions, such as on which compounds to run MD refinement or perform physical experiments. Getting similar estimates from black-box predictive models is more challenging, typically requiring training expensive deep ensembles \citep{rahaman2021uncertainty} or auxiliary confidence models \citep{jumper2021highly, rhodes2025orb}.

\paragraph{Compound generalization} Figure \ref{fig:ahfe_fig}b shows that there is no meaningful correlation between a compound's similarity to the training set (measured by maximum Tanimoto similarity to any compound in the training set) and the resulting AquaGen AHFE absolute error. We hypothesize that since the model is highly regularized by learning to imitate the conformational ensemble induced by classical force-field simulations, the bulk of its capacity is presumably spent modeling fine-grained details that generalize over compound structures.

\paragraph{Error cancellation.}
Inspection of the errors in the first off-diagonal of the MBAR free-energy matrix $\Delta \hat{G}$  (Figure \ref{fig:ahfe_fig}c) shows that AquaGen tends to produce negative $\Delta G$ errors during the electrostatic annihilation stage ($\lambda_k, k \in [1, 5]$) and positive errors during the van der Waals (VDW) annihilation stage ($\lambda_k, k \in [6,20]$). This highlights a limitation of the AE metric (\eqref{eq:ahfe_ae}): since the total free energy difference is computed by summing free-energy increments along the first off-diagonal of $\Delta \hat{G}$, a model can achieve a low AHFE AE despite substantial local errors if those errors cancel across the alchemical pathway. Consequently, AHFE AE alone does not distinguish between models that accurately capture the underlying physics and those that benefit from fortuitous cancellation. To address this issue, we introduce the cumulative absolute error (CAE),
\begin{align}
\label{eq:cae_def}
\text{CAE}
=
\sum_{i=1}^{K-1}
\left|
\Delta \hat{G}^{\text{model}}_{i,i+1}
-
\Delta \hat{G}^{\text{MD}}_{i,i+1}
\right|,
\end{align}
which measures the total deviation along the alchemical pathway and is robust to error cancellation. In \S~\ref{sec:ablations}, we use both AE and CAE to evaluate the impact of various design choices. See \S\ref{app:error_cancellation} for a more detailed discussion of error cancellation.

\begin{table}[t]
\centering
\caption{Comparing the AHFE mean AE of black-box regressors with AquaGen on various held-out test splits which assess both interpolation and extrapolation. Since AquaGen operates at all-atom resolution, we are able to apply MD refinement on the samples. AquaGen (160M) produces samples which, with short 40 ps refinement, achieve < 0.9 kcal/mol AHFE mean AE on various out-of-distribution data splits.}
\label{tab:baselines}
\begin{tabular}{lcccc}
\toprule
\textbf{Model} 
& \makecell{\textbf{AHFE target split} \\ extrapolation \\ $\downarrow$ (kcal/mol)}
& \makecell{\textbf{AHFE target split}\\ interpolation \\ $\downarrow$ (kcal/mol)}
& \makecell{\textbf{FreeSolv} \\ $\downarrow$ (kcal/mol)}
& \makecell{\textbf{CombiSolv} \\ $\downarrow$ (kcal/mol)} \\
\midrule
\makecell[l]{Random Forest \\ (ECFP Fingerprint)} 
& 8.29 {\tiny $\pm 0.82$} & 2.76 {\tiny $\pm 0.66$} & 4.56 {\tiny $\pm 0.27$} & 5.25 {\tiny $\pm 0.30$} \\

GNN (Vacuum Structure) 
& 2.15 {\tiny $\pm 0.36$} & 1.34 {\tiny $\pm 0.34$} & 3.30 {\tiny $\pm 0.28$} & 2.75 {\tiny $\pm 0.26$} \\

GNN (Solvated Structure) 
& 2.05 {\tiny $\pm 0.32$} & 1.00 {\tiny $\pm 0.32$} & 3.23 {\tiny $\pm 0.16$} & 2.60 {\tiny $\pm 0.18$} \\



AquaGen (160M) 
& 2.39 {\tiny $\pm 0.65$} & 1.24  {\tiny $\pm 0.54$} & 3.92 {\tiny $\pm 0.55$} & 3.06 {\tiny $\pm 0.62$} \\

\makecell[l]{AquaGen (160M) \\ w/ MD refinement (40ps)}
& 0.87 {\tiny $\pm 0.26$} & 0.65 {\tiny $\pm 0.25$} & 0.40  {\tiny $\pm 0.19$} & 0.62 {\tiny $\pm 0.17$} \\
\bottomrule
\end{tabular}
\end{table}

\paragraph{Comparison to Black-Box Regressors. } We compare AquaGen to baselines that produce AHFE estimates in a \emph{black-box} fashion, i.e., directly as the output of a regression task. We consider two regression variants: 1) a random forest (RF) model trained on RDKit Extended Connectivity fingerprints (ECFP), a geometry-agnostic representation of the solvated compound, and 2) a GNN trained on the energy-minimized (vacuum or solvated) configuration of the system of interest. We evaluate models on various data splits designed to test interpolation and extrapolation capabilities based on the target AHFE value (see \S\ref{app:appendix_results} for more details). A 160M AquaGen model outperforms RF and performs comparably with the GNN baselines, despite not being explicitly trained to predict AHFE as the baselines were. Due to the refinability of the predictions, we can run very short (40 ps) MD simulations from the generated samples and obtain significantly better AHFE predictions (< 0.9 kcal/mol error on all splits). We also find that relative to AquaGen (Figure \ref{fig:ahfe_fig}b), the vacuum GNN and random forest baselines exhibit a stronger negative correlation between Tanimoto similarity to the training set and AHFE absolute error (Figure \ref{fig:baselines_tan_vs_err}), suggesting slightly poorer generalization. 

\subsection{Ablations}
\label{sec:ablations}
We explore the tradeoff between AE and CAE caused by error cancellation,  by running various ablations. To keep training costs reasonable, we perform all ablations on the 40M parameter base model, rather than the 160M parameter model on which we reported previous results. A summary of all ablations is provided in Table \ref{tab:ablation_ahfe}, with variants to the base run sorted by ascending AHFE AE (we still report the unmodified 160M parameter model results in the Table for completeness). We find that AHFE AE correlates only loosely with Overall CAE, suggesting that error cancellation plays a significant role in the final results. 

\begin{table}[t]
\centering
\caption{Ablation study on model architecture and sampling choices. Variants to AquaGen Base (40M) are sorted by ascending AHFE Mean AE. Due to error cancellation, AHFE AE correlates only weakly with Overall CAE. Some variants (e.g., centering) improve AE, but at the expense of CAE, indicating less physical samples. }
\label{tab:ablation_ahfe}
\begin{tabular}{lcccc}
\toprule
\textbf{Model Variant} 
& \makecell{\textbf{Electrostatic CAE} \\ $\downarrow$ (kcal/mol)}
& \makecell{\textbf{van der Waals CAE} \\ $\downarrow$ (kcal/mol)}
& \makecell{\textbf{Overall CAE} \\ $\downarrow$ (kcal/mol)}
& \makecell{\textbf{AHFE AE} \\ $\downarrow$ (kcal/mol)} \\
\midrule
AquaGen Base (160M) & 2.82 {\tiny $\pm 0.42$} & 2.24 {\tiny $\pm 0.13$} & 5.04 {\tiny $\pm 0.48$} & 1.22 {\tiny $\pm 0.31$} \\
AquaGen Base (160M){\small  + MD (40ps)\hspace{-0.6cm}} & 0.43 {\tiny $\pm 0.14$} & 0.42 {\tiny $\pm 0.12$} & 0.88 {\tiny $\pm 0.26$} & 0.46 {\tiny $\pm 0.15$} \\
\midrule
AquaGen Base (40M)& 3.41 {\tiny $\pm 0.46$} & 1.38 {\tiny $\pm 0.13$} & 4.76 {\tiny $\pm 0.54$} & 2.21 {\tiny $\pm 0.45$} \\
Autoguidance & 2.55 {\tiny $\pm 0.38$} & 2.28 {\tiny $\pm 0.21$} & 4.74 {\tiny $\pm 0.42$} & 1.44 {\tiny $\pm 0.30$} \\
Compound Centering & 5.03 {\tiny $\pm 0.61$} & 4.47 {\tiny $\pm 0.26$} & 9.45 {\tiny $\pm 0.70$} & 1.64 {\tiny $\pm 0.44$} \\
$\mathcal{N}(0, 1)$ prior & 3.36 {\tiny $\pm 0.46$} & 1.70 {\tiny $\pm 0.19$} & 4.99 {\tiny $\pm 0.60$} & 2.50 {\tiny $\pm 0.48$} \\
Auto-$x_0$-variance prior & 5.63 {\tiny $\pm 0.73$} & 1.67 {\tiny $\pm 0.19$} & 7.19 {\tiny $\pm 0.79$} & 4.17 {\tiny $\pm 0.70$} \\
Linear timestep integration & 2.89 {\tiny $\pm 0.80$} & 4.23 {\tiny $\pm 1.00$} & 6.92 {\tiny $\pm 1.58$} & 6.75 {\tiny $\pm 1.63$} \\
\bottomrule
\end{tabular}
\end{table}

\paragraph{Compound centering.} Translating the compound center-of-mass to the center of the simulation box in all MD frames used for training leads to a significant degradation in the CAE of both alchemical legs. We hypothesize that centering the compound creates a strong bias for the model, resulting in an over-repulsion of the solvents around the origin.
While this run achieves better AHFE AE than \textbf{Base (40M)} due to error cancellation, we opt not to use it due to the very high CAE. 

\paragraph{Gaussian prior variance.} The Auto-$x_0$-variance prior is an attempt to reduce the discrepancy between the variance of the prior and target positions $x_0$ and $x_1$. We set the prior simulation box length $\ell$ based on an empirically fitted linear relationship with the square root of number of atoms, $\ell=0.41\sqrt{N} + 8.59$ (such that the prior simulation box itself is $c_0 = \ell I_3$), and the variance of the prior positions $x_0$ such that 95\% of the probability density is contained within $c_0$ prior to periodic wrapping. Surprisingly, this yields worse results across the board, with a particularly negative effect on electrostatic-leg CAE. Setting the prior variance position variance too small (i.e. $\mathcal{N}(0,1)$) creates extreme path crossing and curvature near $t=0$ during training and integration. We find that a better middle ground of $\sigma^2=2$ yields the best results (AquaGen Base). The downside of this choice is that due to the expansion in volume from $x_0$ to $x_1$, we induce an overestimation of the box size. See \S\ref{app:velocity_curvature} for more discussion.

\paragraph{Uniform timestep integration.} Replacing the exponential time discretization (\S \ref{app:velocity_curvature}) with a uniform discretization over $t \in [0,1]$ at inference time leads to a significant degradation in van der Waals CAE, and ultimately a degradation in AHFE AE to 6.75 kcal/mol. This is consistent with our observation that the curvature of the learned marginal velocity field is considerably higher near $t \approx 0$ (Figure~\ref{fig:angular_fig}). We hypothesize that this effect is substantially more pronounced for the van der Waals leg because the model must coordinate the global spatial organization of many solvent molecules extremely early in the integration trajectory, making accurate resolution of the small-$t$ regime particularly important. In contrast, errors in the electrostatic leg appear to arise more locally from imperfect solvent orientation and hydrogen-bond alignment, which accumulate gradually throughout the integration and are less sensitive to the timestep schedule.

\paragraph{Autoguidance.} Autoguidance \citep{karras2024guiding} replaces the unconditional model typically used in classifier-free-guidance \citep{ho2022classifier} with a $\lambda$-conditional model from an earlier checkpoint in training. This yields a significant improvement in Electrostatic CAE from 2.58 to 2.00 kcal/mol, but worsens the van der Waals and overall CAE. Despite an improved AHFE AE of 1.44 kcal/mol from error cancellation, we don't observe similar cancellation when applying autoguidance to the 160M parameter model, so we elect not to use autoguidance for the final results. We believe that this is because the 160M parameter model is already estimating electrostatics fairly accurately, making autoguidance somewhat redundant. However, this points to the potential of exploring guidance strategies to counteract certain model biases, and the importance of improving the expressivity of $\lambda$ conditioning. 

\section{Related Work}
\paragraph{Biomolecular generative models.} 
Folding models~\citep{jumper2021highly, abramson2024accurate} have induced a paradigm shift in protein structure prediction. More recently, generative modeling of atomistic Boltzmann distributions has seen rapid progress through normalizing flows~\citep{noe2019boltzmann, kim2024scalable}, diffusion models~\citep{jing2024generative, lewis2025scalable}, and flow matching approaches~\citep{lipman2023flow,hassan2024flow}. However, many existing approaches operate on reduced-dimensionality manifolds (e.g., torsional angles, internal coordinates, or coarse-grained frames) or on discrete states of a simplified dynamical representation such as a Markov State Model~\citep{kapusniak2025mars}. Although solvent effects are typically incorporated in the underlying training data, to our knowledge, no existing model explicitly includes solvent atoms in its generated configurations. This would be a crucial limitation in our gray-box approach of computing AHFE via potential energy evaluations, since the energetic precision of implicit solvent force fields still lags behind that of explicit solvent ~\citep{tan2006well}, especially for macromolecules~\citep{katkova2017accuracy,robinson2016are}. Hence, explicit water models, such as TIP3P~\citep{price2004tip3p}, remain the standard practice in industrial MD workflows~\citep{schindler2020large,thaler2026boltz}. 
We contrast the characteristics of recent models with this work (AquaGen) in Table \ref{tab:state-of-the-art}.
\begin{table}[h]
    \centering
    \caption{Comparison of recent biomolecular generative models. Our model, AquaGen, is the first to model the Boltzmann conformational ensemble at all-atom resolution, including hydrogens and solvent atoms.}
    \label{tab:state-of-the-art}
    \resizebox{\textwidth}{!}{%
    \begin{tabular}{lcccc}
        \toprule
        \textbf{Model} & \textbf{Output Resolution} & \textbf{Objective} & \textbf{Generates Solvent} & \textbf{Systems} \\
        \midrule
        Scalable Flows~\citep{kim2024scalable} 
        & All heavy-atoms
        & Conformational ensemble 
        & No 
        & Proteins \\

        ESM-Flow~\citep{jing2024alphafold} 
        & $\beta$-Carbon atoms 
        & Conformational ensemble 
        & No 
        & Proteins \\

        AlphaFold 3~\citep{abramson2024accurate} 
        & All heavy atoms  
        & Structure prediction 
        & No 
        & Proteins, ligands, nucleic acids \\

        Bio-Emu~\citep{lewis2025scalable}
        & Backbone heavy atoms 
        & Conformational ensemble 
        & No 
        & Proteins \\

        Boltz-2~\citep{passaro2025boltz} 
        & All heavy atoms 
        & Structure + affinity prediction
        & No 
        & Proteins, ligands, nucleic acids \\

        MarS-FM~\citep{kapusniak2025mars} 
        & All heavy atoms
        & Conformational ensemble
        & No 
        & Proteins \\

        ATMOS~\citep{shi2026atomic} 
        & All heavy atoms 
        & Dynamics 
        & No 
        & Proteins, protein--ligand complexes \\
        
        \midrule
        \textbf{AquaGen (this work)}
        & \textbf{All-atom} 
        & \textbf{Conformational ensemble} 
        & \textbf{Yes} 
        & \textbf{Solvated compounds} \\
        \bottomrule
    \end{tabular}%
    }
\end{table}

\paragraph{Traditional free energy methods.}
Free energy estimation is a classical problem in molecular simulation, with a long history of methods including free energy perturbation (FEP)~\citep{zwanzig1954high}, thermodynamic integration (TI)~\citep{kirkwood1935statistical,frenkel2001understanding}, umbrella sampling~\citep{torrie1977nonphysical}, replica exchange~\citep{hukushima1996exchange}, Bennett acceptance ratio (BAR)~\citep{bennett1976efficient}, and multistate Bennett acceptance ratio (MBAR)~\citep{shirts2008statistically}. These approaches are statistically principled and remain standard in computational chemistry, but they rely on extensive MD sampling, often across many intermediate alchemical or thermodynamic states. As a result, their practical cost is dominated by the need to generate sufficiently decorrelated samples with adequate phase-space overlap between neighboring states.

\paragraph{Learned free energy estimation.}
A separate line of work has explored using generative models and learned transport maps to accelerate free energy calculations. Neural Thermodynamic Integration~\citep{mate2024neural} replaces a hand-designed alchemical path with a trainable neural Hamiltonian, enabling thermodynamic integration through learned intermediate ensembles, with follow-up work applying this idea to solvation free energy estimation~\citep{mate2025solvation}. FEAT~\citep{he2025feat} is another complementary direction, using adaptive learned transports to construct free energy estimators based on non-equilibrium identities such as the escorted Jarzynski equality and controlled Crooks relations. In general, the scalability of these methods has not been demonstrated beyond low-dimensional settings. Recent biomolecular models such as BioEmu~\citep{lewis2025scalable}, AlphaFold3~\citep{abramson2024accurate}, and Boltz-2~\citep{passaro2025boltz} have also been extended toward conformational ensemble generation and binding affinity prediction. However, these approaches generally do not produce explicit-solvent Boltzmann samples suitable for direct potential-energy evaluation and MBAR reweighting.

\section{Conclusions}

In this work, we introduced AquaGen, an atomistic generative model that efficiently samples from the Boltzmann distribution of solvated drug compounds, in the all-atom ($\sim10^3$ atoms), explicit-solvent, PBC-aware setting. This enables the computation of absolute hydration free energy in a \emph{gray-box} fashion. Although the samples are drawn from a \emph{black-box} ML model, they are inspectable, and their energies are computed with physical force fields, making the ensemble AHFE predictions of this framework grounded in the underlying physics. While there is still room for improvement, particularly in the expressivity of $\lambda$ conditioning and the accuracy of box volume estimation, we have demonstrated benefits from self-consistent error cancellation and train- and test-time computation to provide accurate free energy estimates, and the ability to easily extract calibrated uncertainties from the model.

\paragraph{Future work.} In future work, we hope to further demonstrate the benefits of our gray-box approach and of modeling biophysical problems at a level of resolution immediately compatible with potential energies used in standard MD. To this end, we plan to improve our modeling of rigid water and improve the expressivity of conditioning on the alchemical parameter $\lambda$. Finetuning or test-time search strategies \citep{domingo2024adjoint, singhal2025general} aimed at regions of high distributional overlap or towards alignment with experimentally obtained free energies \citep{lewis2025scalable} could also be fruitful directions. Ultimately, we hope to leverage the proposed framework to tackle more problems for which MD is a classical solution in real-world industrial settings. Solving problems such as the prediction of lipophilicity (logP), membrane permeability, or absolute binding free energy (ABFE) --which are significantly more challenging due to their scale, system heterogeneity, and pipeline complexity--would be transformative for drug discovery.

\paragraph{Acknowledgements.} We would like to thank Daniel Cutting for his contributions to an earlier instantiation of the AquaGen model. We would also like to thank previous Valence interns Rokas Elijosius, John Gardner, Siddarth Venkatraman, and Vineet Jain for many helpful discussions and suggestions throughout the development of AquaGen. Finally, we are grateful to Hatem Helal, Gerhard Koenig, Geoff Wood, William Wu, and Will Glass for their valuable feedback during the course of this project.


\newpage

\bibliography{main}
\bibliographystyle{valence}

\newpage

\appendix
\section{Model Details}

\subsection{Model Architecture}
\label{app:arch}

We parameterize the flow-matching velocity field $v_\theta(\bar{x}_\tau, \tau | \lambda)$ with a graph neural network (GNN) operating on the joint system state $\bar{x}_\tau = \{x_\tau, c_\tau\}$, where $x_\tau \in \mathbb{R}^{N \times 3}$ denotes atomic coordinates and $c_\tau$ denotes the periodic simulation cell represented via virtual nodes. Given atomic coordinates $x_\tau$, we construct a graph
\begin{align}
    \mathcal{G}
    =
    (\mathcal{V},
    \mathcal{E}_{\text{bond}}
    \cup
    \mathcal{E}_{\text{dist}}),
\end{align}
where $\mathcal{E}_{\text{bond}}$ contains covalent bond edges and $\mathcal{E}_{\text{dist}}$ contains geometric $k$-nearest-neighbor edges computed under periodic boundary conditions using the minimum-image convention. Node features include standard chemical descriptors such as atom type, charge, and force-field features. Edge features include bond order and relative positional information.
The model operates on relative displacement vectors
\begin{align}
    \Delta x_{ij}
    =
    x_{\tau,j} - x_{\tau,i},
\end{align}
rather than absolute coordinates. 
The network is conditioned on both the flow-matching time $\tau$ and alchemical parameter $\lambda$.

\paragraph{Periodic boundary condition representation.}
To represent the periodic simulation cell $c_\tau$, we introduce virtual nodes corresponding to the shape 
of the simulation box, which together fully specify the box geometry. These virtual nodes' 
velocities are jointly predicted as part of the learned vector field $v_\theta(\bar{x}_\tau, \tau | \lambda)$. 

In the Gaussian prior distribution $p_0(\bar{x})$, the virtual node coordinates are initialized such that the corresponding simulation box contains approximately $95\%$ of the atomic coordinates sampled from the coordinate prior. For our chosen coordinate prior $x_0 \sim \mathcal{N}(0, 2I_{3N})$ , this corresponds to a mean box length of $\mu_c = 7.84 $ and a variance of $\sigma^2_c = 0.05$. 

\paragraph{Water compression.}

To reduce the cost of message passing in solvent-dominated systems, we compress each water molecule into a single latent node during the GNN backbone. We retain the oxygen atom as the representative message-passing node and remove the two associated hydrogen nodes from the graph. The local water geometry is encoded, %
using the two O--H displacement vectors. 
and the resulting feature $h_{\text{water}}$ is attached as an additional latent feature to the retained oxygen node. Message passing is then performed on the compressed graph, in which non-water atoms and water oxygens remain explicit, but water hydrogens are omitted.

Let $z_i^{\mathrm{cmp}}$ denote the output latent representation produced by the message-passing backbone for node $i$ in the compressed graph. For a retained water oxygen $O$, the corresponding compressed-graph output $z_O^{\mathrm{cmp}}$ is used both as the oxygen representation and to reconstruct latent representations for the two omitted hydrogens. In particular, we apply learned projection networks to produce
\begin{align}
\left(
z_{H_1}^{\mathrm{exp}},
z_{H_2}^{\mathrm{exp}}
\right)
=
\phi_{\mathrm{proj}}
\left(
z_O^{\mathrm{cmp}}
\right),
\end{align}
and set $z_O^{\mathrm{exp}} = z_O^{\mathrm{cmp}}$. Here $z_i^{\mathrm{exp}}$ denotes the reconstructed explicit-atom latent representation after decompressing the water molecules. The reconstructed hydrogen representations are then inserted back into the original atom ordering, yielding an explicit latent representation ${z^{\mathrm{exp}}} \in \mathbb{N \times d}$ over all $N$ atoms. The prediction head is applied to these latent features to produce per-atom velocities.

Since the majority of atoms in solvated systems are water hydrogens, this reduces the number of solvent nodes passed through the backbone from three per water molecule to one per water molecule. This corresponds to an overall FLOP reduction of roughly $\frac{2}{3}$ in water-dominated systems, with minimal loss in accuracy for sufficiently large models.



\paragraph{Model scaling.}

We generally find that scaling model width, depth, and message-passing horizon produce comparable improvements, provided that nodes have sufficient receptive field. In practice, increasing width is typically the most computationally efficient scaling direction.

\section{Molecular Dynamics Dataset}
\label{app:appendix_data}

Internal molecules in the training dataset were selected as products of medicinal-chemistry lead optimization, and only uncharged compounds were retained to avoid complications associated with ionization-state and finite-size corrections. To ensure methodological consistency across datasets, the reference AHFE values for both FreeSolv and CombiSolv were recomputed using the same protocol applied to the internal compounds.

All calculations were performed with the OpenFE absolute solvation workflow, which implements an absolute hydration free energy thermodynamic cycle in which the ligand is transformed in both solvent and vacuum. In this protocol, electrostatic interactions are fully annihilated first, followed by Lennard-Jones decoupling with intramolecular Lennard-Jones interactions preserved while removing intermolecular van der Waals interactions using Gapsys's soft-core potential. Amino acids were parameterized using AMBER ff14SB force field. Small molecules were parameterized using the OpenFF 2.1.1 force field with AM1-BCC charges. For the solvent leg, we used the TIP3P water model. We used a Middle Langevin integrator and performed equilibration and production in the NPT ensemble with a Monte Carlo barostat.

The alchemical path was discretized into 20 $\lambda$ values with schedules:
\begin{align}
\lambda_\text{elec} &= [0.0, 0.25, 0.5, 0.75, 1.0, 1.0, 1.0, 1.0, 1.0, 1.0, 1.0, 1.0, 1.0, 1.0, 1.0, 1.0, 1.0, 1.0, 1.0, 1.0] \\
\lambda_\text{vdW} &= [0, 0, 0, 0, 0, 0.05, 0.1, 0.2, 0.3, 0.4, 0.5, 0.6, 0.65, 0.7, 0.75, 0.8, 0.85, 0.9, 0.95, 1.0]
\end{align}

Electrostatics were turned off over the first five windows, after which van der Waals interactions were gradually decoupled across the remaining windows, consistent with the OpenFE AHFE strategy. Sampling was enhanced using Hamiltonian Replica Exchange (HREX). For each compound, we ran 3 independent replicas, each for 10 ns, using HREX at 298.15 K and 1 bar. A 4 fs timestep was used, enabled by hydrogen mass repartitioning (3.0 amu). Using a sampling frequency of 1 ps, 10,000 frames were extracted from production trajectories across each compound, replica, and $\lambda$ window. We trained on over 2000 compounds, yielding approximately over 1 billion training frames. Hydration free energies, \(\Delta G_\text{AHFE}\), were obtained by applying the MBAR estimator as implemented in PyMBAR 4.0. 

\section{Additional Results and Details}
\label{app:appendix_results}
\subsection{Curvature of learned velocity field.} 
\label{app:velocity_curvature}

We find that our choice of Gaussian prior leads to non-uniform curvature of the learned velocity field over the flow matching time interval $\tau \in [0,1]$. As a proxy for curvature, we measure the discretized \emph{angular velocity} between two adjacent timesteps during sampling via the arccos of the cosine similarity between consecutive $v_\theta(x, \tau|\lambda)$:
\begin{align}
    v_t &= v_\theta(x_t, t| \lambda)\\
    \omega_\tau &= \frac{ v_\tau \cdot v_{\tau+\Delta \tau}}{\|v_\tau\|\|v_{\tau+\Delta \tau}\|\Delta \tau}
\end{align}

\begin{figure}[H]
    \centering
    \ifvtwofig
    \includegraphics[width=0.45\linewidth]{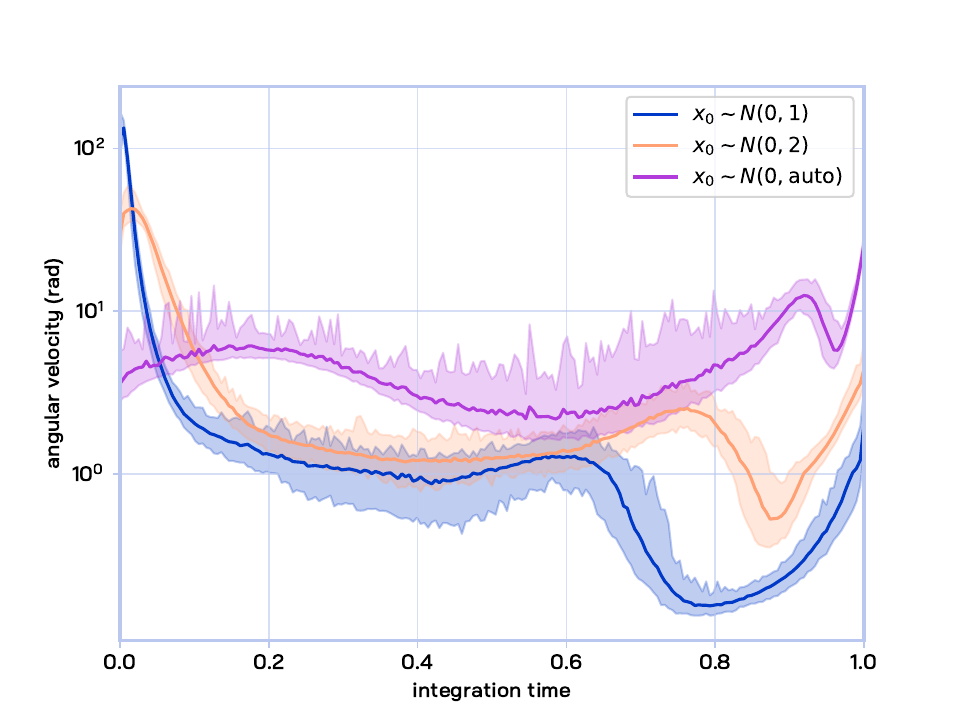}
    \else
    \includegraphics[width=0.45\linewidth]{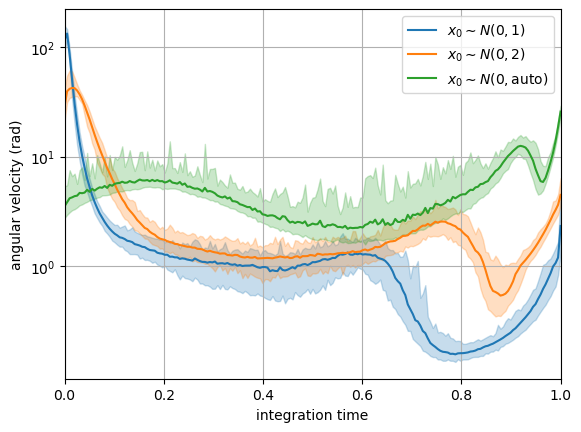}
    \fi
    \caption{\textbf{Angular velocity of $v_\theta(x, \tau|\lambda)$} during integration. The large velocity near $t=0$ can be counteracted by an exponential integration schedule for the $N(0,1)$ and $N(0,2)$ models. The auto-variance prior reduces the angular velocity near $t=0$, but leads to higher angular velocities near $t=1$ and higher overall CAE values.}
    \label{fig:angular_fig}
\end{figure}

As shown in Figure \ref{fig:angular_fig}, the angular velocity is considerably higher near $t=0$ for the Gaussian priors with a fixed variance of either 1 or 2. We hypothesize that this is due to the change in scale from the prior to the target distribution. This motivates the use of an exponential timestep schedule during sampling which employs a finer discretization near $t=0$. Specifically, we define a normalized auxiliary variable $u \in [0,1]$ with uniform spacing and map it to the integration schedule
\begin{align}
    t(u)
    =
    \frac{\exp(\alpha u) - 1}
    {\exp(\alpha) - 1},
\end{align}
where $\alpha > 0$ controls the concentration of timesteps near $t=0$. Larger values of $\alpha$ allocate more integration steps near the prior distribution, where the learned vector field exhibits higher curvature. In all experiments, we use $\alpha = 4$.

The auto-variance prior has lower angular velocity at $t=0$ but higher angular velocity at $t=1$. We tried counteracting this with various integration schedules (e.g., a reverse exponential which allocates more timesteps near $t=1$), but were unable to achieve competitive results with that of the base model which uses a Gaussian prior with a fixed variance of 2.

\subsection{Additional tICA plots.}
\label{app:additional_tica}

\begin{figure}[H]
    \centering
    \ifvtwofig
    \includegraphics[width=0.8\linewidth]{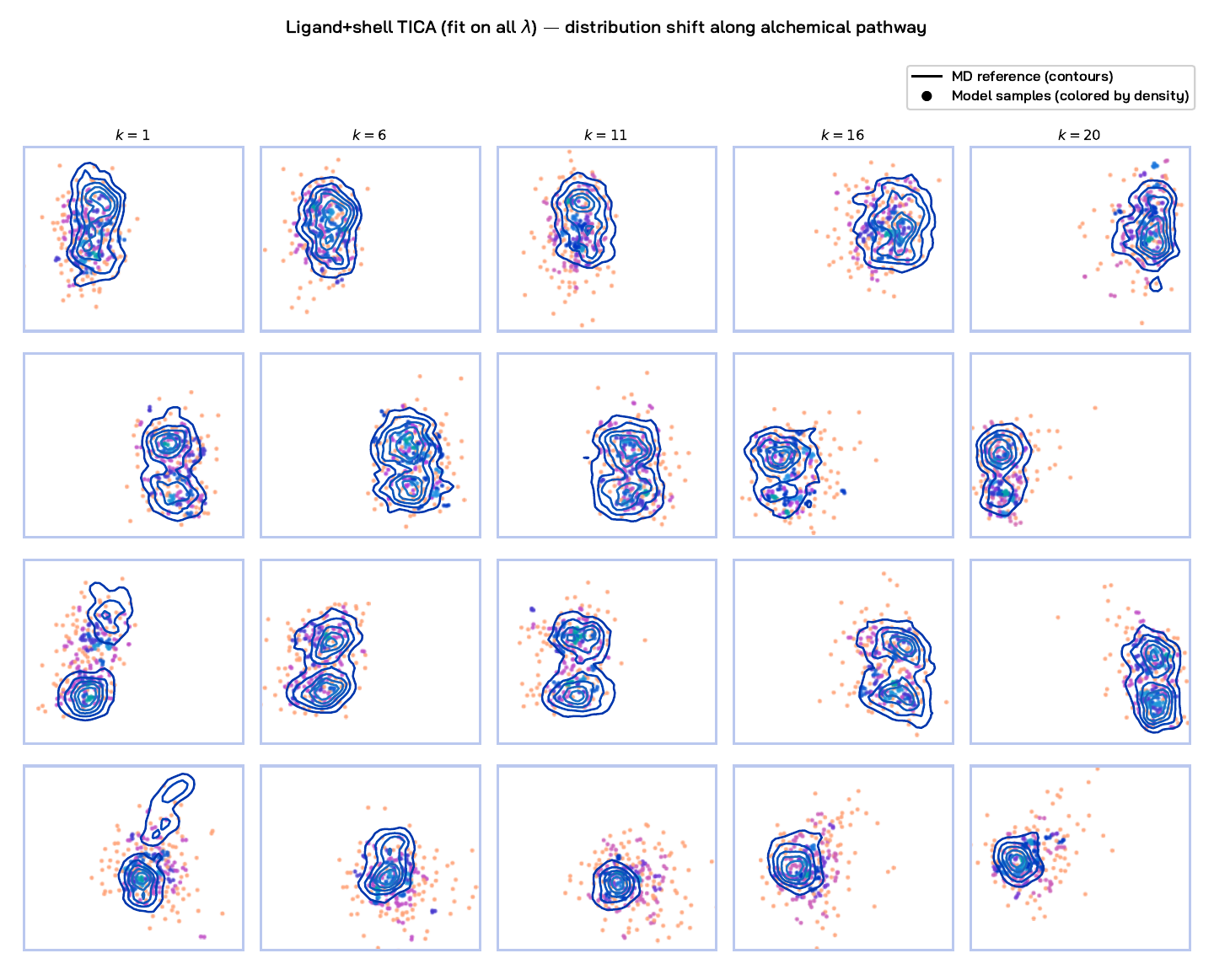}
    \else
    \includegraphics[width=0.8\linewidth]{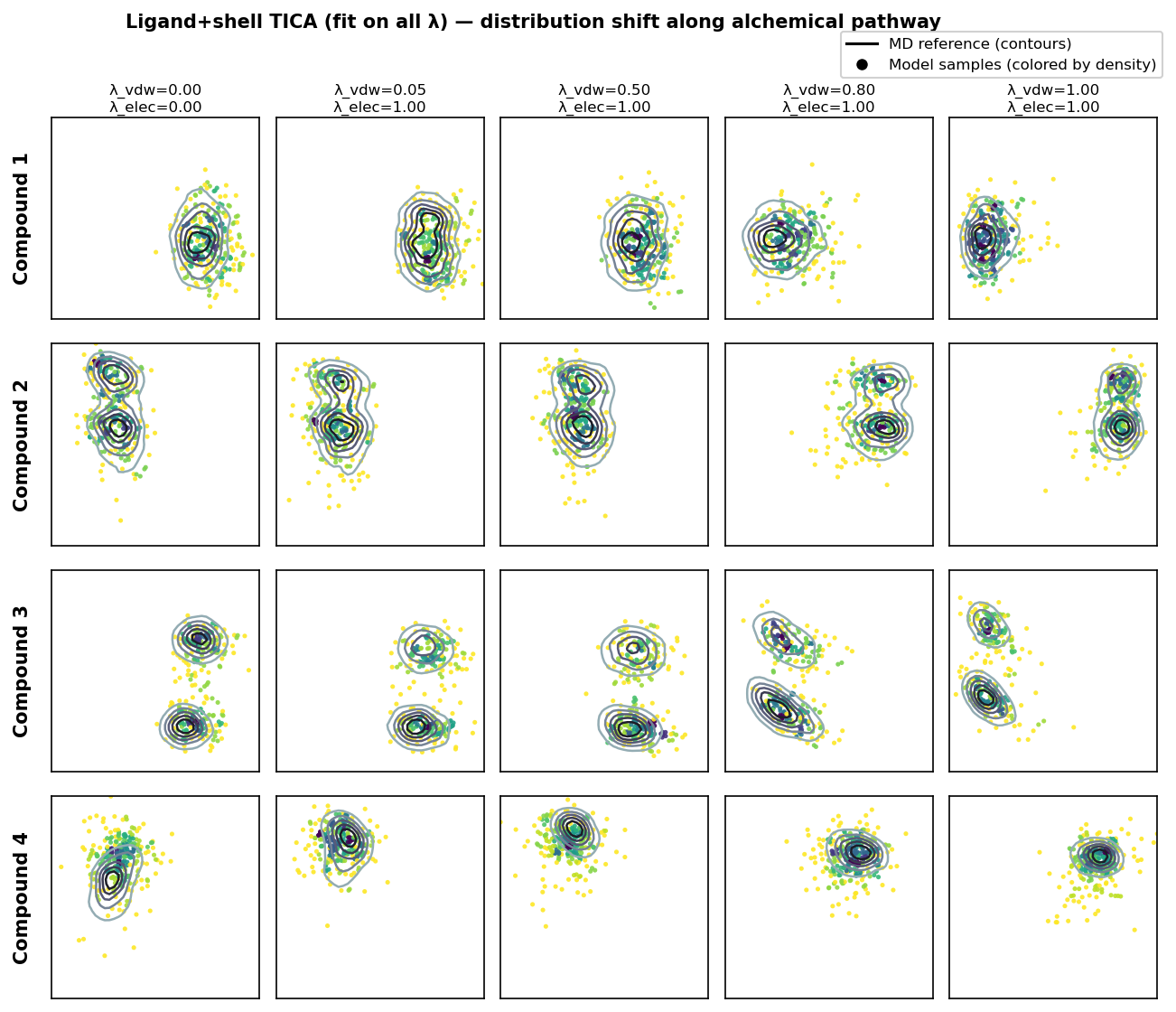}
    \fi
    \caption{\textbf{tICA decomposition} over the alchemical trajectories. AquaGen samples track the change in principal components as $\lambda$ evolves.}
    \label{fig:tica_fig}
\end{figure}

We show tICA plots for more randomly selected compounds from the test set in Figure \ref{fig:tica_fig}. The reference tICA is fit on MD samples across all $\lambda$ values, using as features all pairwise Euclidean distances (in Å) among the ligand heavy atoms, in addition to the 20 water oxygens nearest to the ligand.

\subsection{Additional energy decompositions at fully-interacting distribution.}

\begin{figure}[H]
    \centering
    \ifvtwofig
    \includegraphics[width=0.9\linewidth]{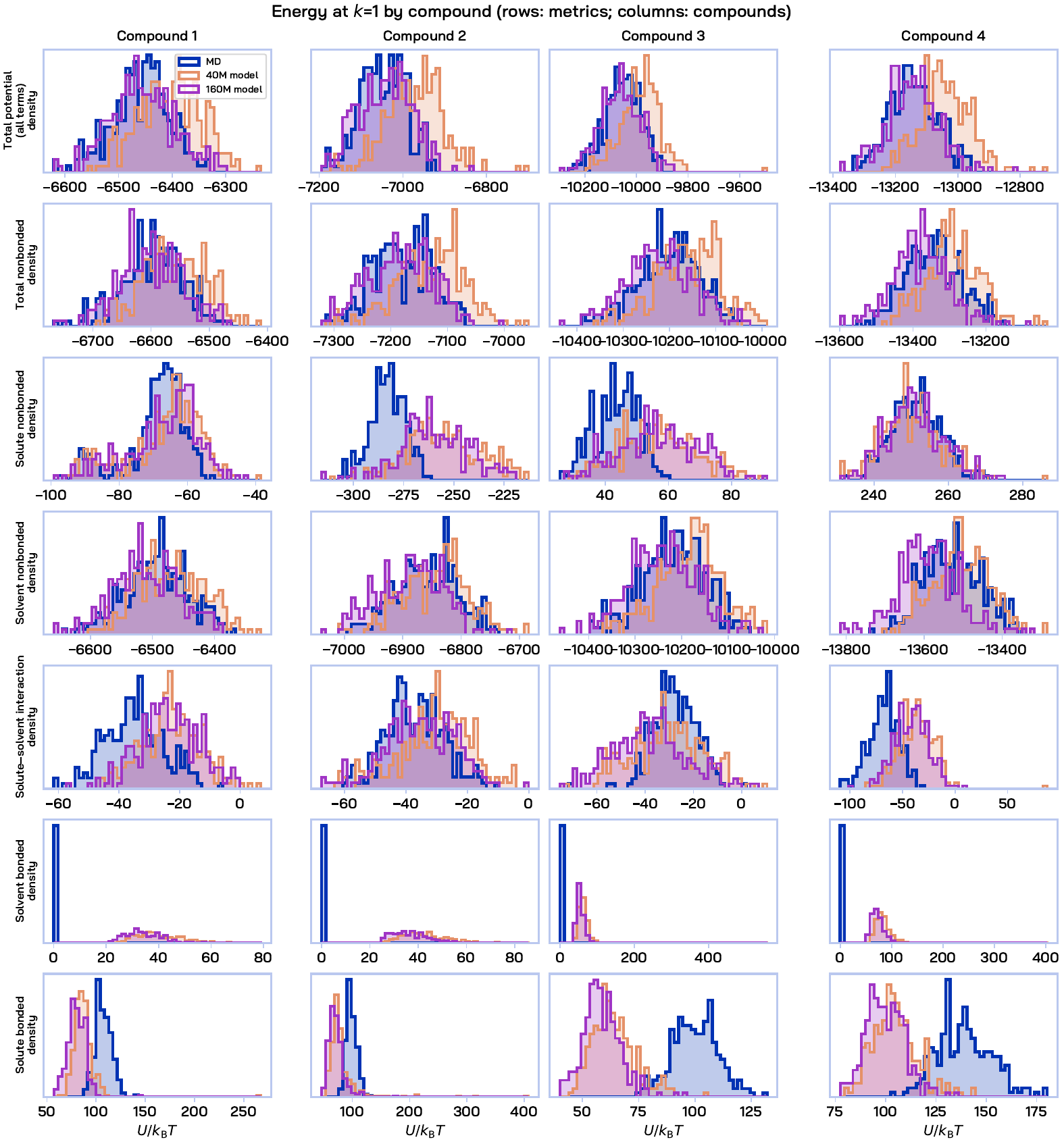}
    \else
    \includegraphics[width=0.9\linewidth]{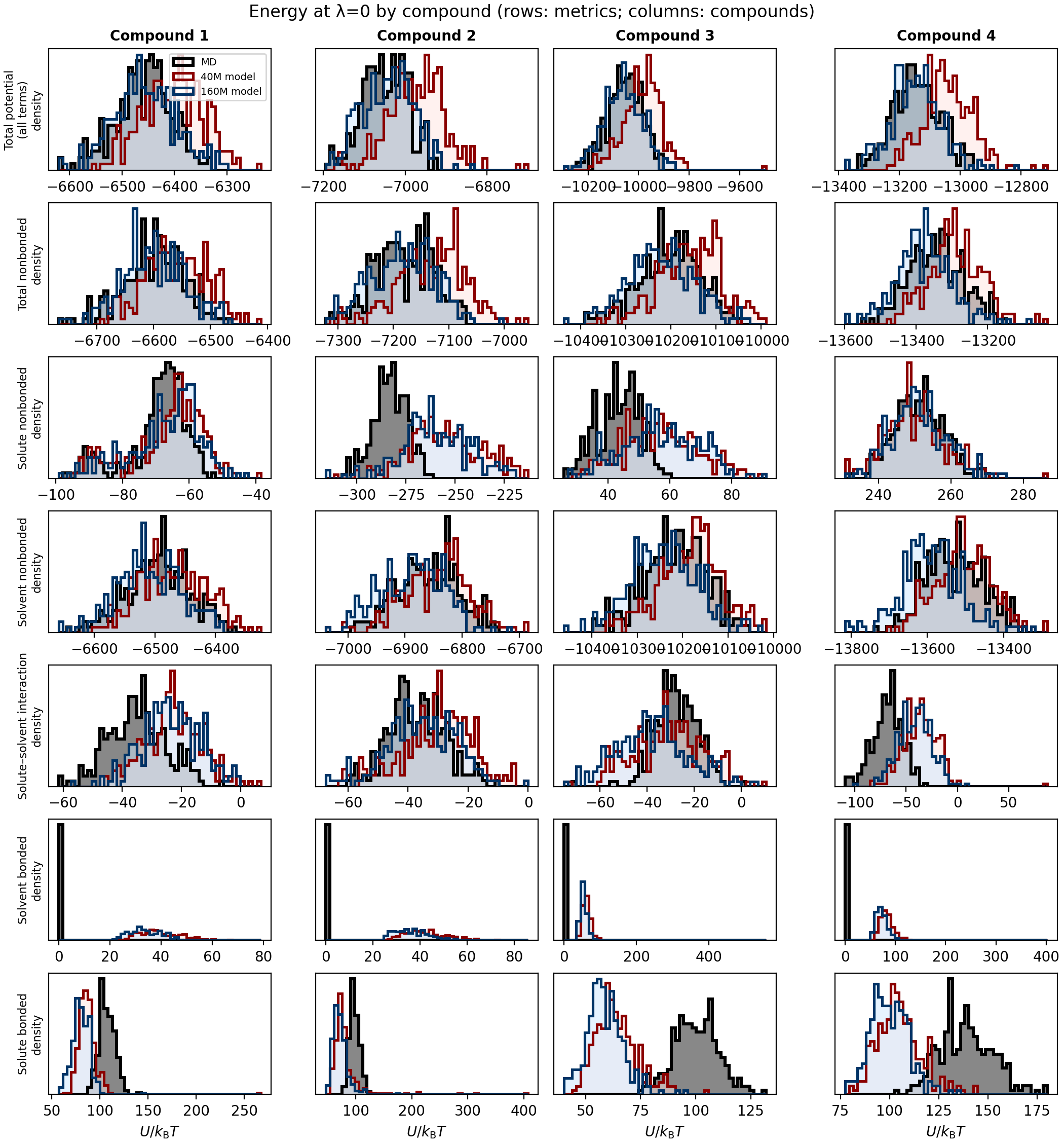}
    \fi
    \caption{\textbf{Energy decompositions at the fully-interacting endpoint ($\lambda_k, k=1$)}. For each compound and energy component, we plot the reference MD distribution (gray), the distribution from the 40M AquaGen model (red), and the distributon from the 160M AquaGen model (blue).}
    \label{fig:decomposition_fig}
\end{figure}

In Figure \ref{fig:decomposition_fig}, we show more granular energy decompositions of samples generated by a 40M and 160M parameter AquaGen model trained only the fully-interacting endpoint ($\lambda_k, k = 1$). Due to the lack of a rigid water assumption in our generative models, we overestimate the solvent bonded energy contribution (second row from bottom) relative to the reference MD samples. We also tend to underestimate the solute bonded contribution, suggesting minor mode collapse which causes excessive stiffness around average bond lengths. In general, scaling the model from 40M to 160M parameters has the largest positive effect on solvent nonbonded energies, which are not modulated by $\lambda$. This suggests that future work should prioritize allocating model capacity towards parts of the system which are more consequential for the downstream application, i.e., AHFE estimation.

\subsection{Error cancellation along the alchemical pathway}
\label{app:error_cancellation}

The free energy difference between the alchemical endpoints can be written as a sum over the first off-diagonal of the MBAR free-energy matrix:
\begin{align}
\Delta \hat{G}_{0,K}
=
\sum_{i=1}^{K-1}
\Delta \hat{G}_{i,i+1}.
\end{align}

Figure \ref{fig:ahfe_fig}c provides additional insight into model behavior along this pathway. The region to the left of the dashed line corresponds to electrostatic annihilation ($\lambda_k, k \in [1, 5]$), while the region to the right corresponds to van der Waals annihilation ($\lambda_k, k \in [6, 20]$). We observe a consistent trend in which the model produces negative $\Delta G$ errors in the electrostatic regime and positive $\Delta G$ errors in the VDW regime.

One possible explanation is the structure of the alchemical schedule. Electrostatic interactions are annihilated over only four windows, whereas VDW interactions are annihilated over fifteen windows. Consequently, adjacent Boltzmann distributions differ more strongly in the electrostatic regime. The observed pattern suggests that the model learns a smoother, averaged evolution across $\lambda$, effectively underestimating the rate of change in the electrostatic region and overestimating it in the VDW region.

The AHFE absolute error can be written as
\begin{align}
\text{AHFE AE}
&=
\left|
\Delta \hat{G}^{\text{model}}_{1,K}
-
\Delta \hat{G}^{\text{MD}}_{1,K}
\right| \\
&=
\left|
\sum_{i=1}^{K-1}
\left(
\Delta \hat{G}^{\text{model}}_{i,i+1}
-
\Delta \hat{G}^{\text{MD}}_{i,i+1}
\right)
\right|.
\end{align}

Under this interpretation, the signed area under the error curve in Figure \ref{fig:ahfe_fig}c corresponds to the final AHFE error. Positive and negative deviations can therefore cancel, making it possible for a model with large local free-energy errors to nevertheless achieve a small AHFE AE.

To quantify these local deviations directly, we define the cumulative absolute error (CAE):
\begin{align}
\text{CAE}
=
\sum_{i=1}^{K-1}
\left|
\Delta \hat{G}^{\text{model}}_{i,i+1}
-
\Delta \hat{G}^{\text{MD}}_{i,i+1}
\right|.
\end{align}

By the triangle inequality,
\begin{align}
\text{AHFE AE}
\le
\text{CAE},
\end{align}
with equality only when all pairwise error terms have the same sign. CAE therefore provides an upper bound on AE that is insensitive to error cancellation and more directly measures inaccuracies in the modeled overlap between adjacent alchemical distributions.

\subsection{Regression baselines.}
We provide additional details on the regression baselines to which we compare AquaGen in Table \ref{tab:baselines}:

\begin{enumerate}
\item \textbf{Random Forest (ECFP Fingerprint):} As a simple, structure-agnostic baseline, we train a random forest model to regress to the ground truth AHFE value given the Extended Connectivity Fingerprint (ECFP) descriptor \citep{ecfp} of the ligand. 
\item \textbf{GNN:} We train a GNN (with the same architecture as AquaGen) to map from the minimal energy configuration of each ligand to the ground truth AHFE value. We consider two variants: 1) providing only the ligand structure (\textbf{GNN (Vacuum Structure)}) and 2) providing both the ligand and water structure (\textbf{GNN (Solvated Structure)}). This baseline is in the spirit of recent works which integrate binding affinity prediction as an auxiliary regression task along with structure prediction \citep{passaro2025boltz}. 
\end{enumerate}

Baselines are trained on all compounds from the AquaGen training set, excluding compounds with true AHFE values in the bottom or top 5\% of the original training data (for fair comparison, the AquaGen models reported in Table \ref{tab:baselines} are also re-trained on the same compounds). All methods are then evaluated on four test splits:
\begin{enumerate}
\item \textbf{Target split of internal dataset - interpolation (100 compounds):} Compounds in the test set whose true AHFE lies within the middle 90\% of values in the original AquaGen training set .
\item \textbf{Target split of internal dataset- extrapolation (224 compounds):} Compounds in the test set whose true AHFE lies in the bottom or top 5\% of values in the original AquaGen training set .
\item \textbf{FreeSolv \citep{mobley2014freesolv}}  (\textbf{589 compounds}): A database of neutral fragment-like compounds, spanning a wide range of molecular weights (around 16 - 499 Daltons) and polarities.
\item \textbf{CombiSolv \citep{vermeire2021transfer} (575 compounds)}: A multi-solvent solvation energy mega-database that was curated from multiple sources, including FreeSolv \citep{mobley2014freesolv}, MNSol \citep{marenich2020minnesota}, CompSol \citep{moine2017estimation}, and the Abraham dataset \citep{grubbs2010mathematical}. Here we used only compounds solvated with water, and compounds originating from FreeSolv were excluded .

\end{enumerate}

Unlike our internal dataset, which is composed of drug-like molecules (median molecular weight of >300 g/mol, median heavy atom count >20), FreeSolv and CombiSolv are mostly comprised of fragment-sized, low-molecular-weight organic solutes (median MW of 120.6 g/mol and 184.3 g/mol, respectively) (relevant statistics are summarized in Table \ref{tab:dataset_stats}).

\begin{table}[H]
\centering
\caption{Comparison of molecular size statistics across the FreeSolv, CombiSolv, and internal dataset.}
\label{tab:dataset_stats}
\begin{tabular}{lccc}
\toprule
\textbf{Metric} &
\textbf{FreeSolv (FS)} &
\textbf{CombiSolv (CS)} &
\textbf{Internal Dataset}\\
\midrule
Total Compounds         & 589& 575   & thousands \\
Median MW (g/mol)       & 120.6 & 184.3 & >300 \\
Median Heavy Atom Count      & 8.0   & 13.0  & >20 \\
95th Percentile MW      & 295.0 & 362.5 & >500 \\
\bottomrule
\end{tabular}
\end{table}

\begin{figure}[H]
    \centering
    \ifvtwofig
    \includegraphics[width=0.95\linewidth]{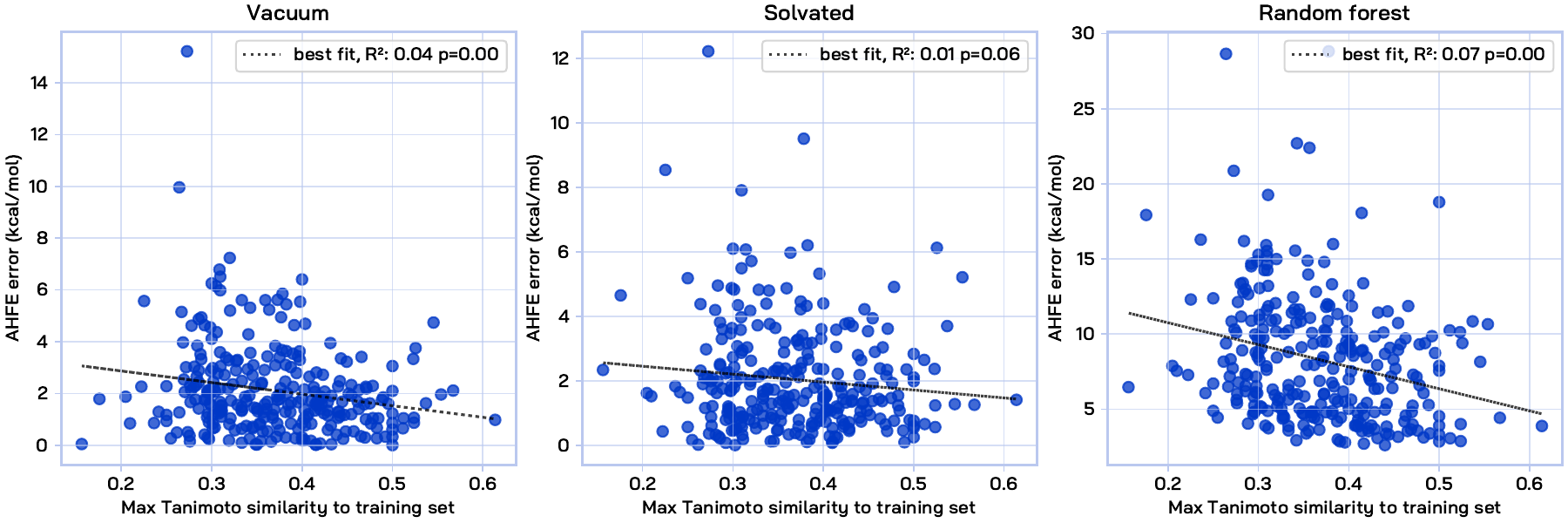}
    \else
    \fi
    \caption{\textbf{Baseline AHFE Mean AE vs similarity to the training set for baselines}. The weak, negative trend suggests that baselines succeed closer to their training set.}
    \label{fig:baselines_tan_vs_err}
\end{figure}

Figure \ref{fig:baselines_tan_vs_err} plots the correlation between AHFE Mean AE and Tanimoto similarity for the regression baselines. The baselines exhibit a stronger negative correlation than AquaGen (Figure \ref{fig:ahfe_fig}b), suggesting they may be less reliable in out-of-distribution settings.

\end{document}